\documentclass[%
reprint,
superscriptaddress,
 amsmath,amssymb,
 aps,
pra,
]{revtex4-2}

\usepackage{amsmath,xparse}
\usepackage{graphicx}
\usepackage{dcolumn}
\usepackage{bm}
\usepackage{xcolor}
\definecolor{mypink}{rgb}{1, 0.2, 0.745}
\definecolor{mycyan}{rgb}{0.090, 0.667, 0.553}
\definecolor{myviolet}{rgb}{0.256, 0.207, 1.00}
\definecolor{freshgreen}{rgb}{0.051, 0.796, 0.733}
\definecolor{mygreen}{rgb}{0.1, 0.72, 0.1}
\definecolor{myviolet}{rgb}{0.256, 0.207, 1.00}
\definecolor{myorange}{rgb}{0.9, 0.44, 0.0}
\definecolor{mypink2}{rgb}{0.898, 0.0, 0.451}

\definecolor{reda}{rgb}{1, 0.123, 0.404}
\definecolor{purplea}{rgb}{0.663, 0.435, 1.0} 
\definecolor{bluea}{rgb}{0.207, 0.259, 1.0}  

\usepackage{hyperref}
\hypersetup{colorlinks, citecolor=mypink,
 linkcolor=myviolet,
 urlcolor=mycyan}

\usepackage{booktabs}
\usepackage[super]{nth}   
\usepackage{lipsum}
\usepackage{todonotes}
\usepackage[normalem]{ulem}

\newcommand\hlineadj[1]{\noalign{\hrule height #1}}

\let\latexchi\chi
\makeatletter
\renewcommand\chi{\@ifnextchar_\sub@chi\latexchi}
\newcommand{\sub@chi}[2]{
  \@ifnextchar^{\subsup@chi{#2}}{\latexchi^{}_{#2}}%
}
\newcommand{\subsup@chi}[3]{
  \latexchi_{#1}^{#3}%
}
\makeatother

\def\REDLINE{0} 
\if\REDLINE1
\newcommand{\rladd}[1]{\textcolor{red}{#1}}
\newcommand{\rlremove}[1]{\sout{#1}}
\else
\newcommand{\rladd}[1]{#1}
\newcommand{\rlremove}[1]{\unskip}
\fi

\def\REDLINETWO{0} 
\if\REDLINETWO1
\newcommand{\rladdtwo}[1]{\textcolor{red}{#1}}
\newcommand{\rlremovetwo}[1]{\sout{#1}}
\else
\newcommand{\rladdtwo}[1]{#1}
\newcommand{\rlremovetwo}[1]{\unskip}
\fi

\begin{document}

\preprint{APS/123-QED}

\title{All-Electron Molecular Tunnel Ionization Based on the \\
Weak-Field Asymptotic Theory in the Integral Representation
}

\author{Imam S. Wahyutama}
\email{iwahyutama@ucmerced.edu}
\affiliation{Department of Physics and Astronomy, Louisiana State University, Baton Rouge, Louisiana 70803, USA}

\author{Denawakage D. Jayasinghe}
\affiliation{Department of Chemistry, Louisiana State University, Baton Rouge, Louisiana 70803, USA}

\author{Fran\c{c}ois Mauger}
\affiliation{Department of Physics and Astronomy, Louisiana State University, Baton Rouge, Louisiana 70803, USA}

\author{Kenneth Lopata}
\affiliation{Department of Chemistry, Louisiana State University, Baton Rouge, Louisiana 70803, USA}
\affiliation{Center for Computation and Technology, Louisiana State University, Baton Rouge, Louisiana 70808, USA}

\author{Kenneth J. Schafer}
\affiliation{Department of Physics and Astronomy, Louisiana State University, Baton Rouge, Louisiana 70803, USA}

\date{\today}

\begin{abstract}

Tunnel ionization (TI) underlies many important ultrafast
processes, such as high-harmonic generation and strong-field ionization. Among the existing theories for TI, many-electron weak-field asymptotic theory (ME-WFAT) is by design
capable of accurately treating many-electron effects in TI. An earlier version of ME-WFAT relied on an accurate representation of the asymptotic tail of the orbitals, which hindered its implementation in Gaussian-basis-set-based quantum chemistry programs. In this work, we reformulate ME-WFAT in the integral representation, which makes the quality of the asymptotic tail much less critical, hence greatly facilitating its implementation in standard quantum chemistry packages. 
The integral reformulation introduced here is therefore much more robust when applied to molecules with arbitrary geometry. 
We present several case studies,
among which is the CO molecule where some 
earlier theories disagree with experiments. Here, we find that ME-WFAT produces the largest ionization probability when the field points from C to O, as experiments suggest. 
An attractive feature of ME-WFAT is that it can be used with various types of multielectron methods whether of density functional [\href{https://doi.org/10.1103/PhysRevA.106.052211}{Phys. Rev. A \textbf{106}, 052211 (2022)}] or multiconfiguration types, \rlremove{which } \rladd{this in turn} facilitates tunnel ionization calculation in systems exhibiting a strong multireference character\rlremove{, such as transition-metal-containing molecules}.
\end{abstract}

\maketitle


\section{Introduction} \label{sec:intro}
In the investigation of strong-field ionization (SFI) \cite{SFI_anatomy, SFI_keldysh, SFI_revisited} in many-electron systems, theoretical methods can provide an in-depth scrutiny that can both guide and illuminate experimental studies \cite{SFI_complexmol}.
In this regard, an accurate and efficient theoretical description of tunnel ionization (TI), which is the dominant method of SFI at long wavelengths, is  desirable, given the time-consuming nature of many-electron simulations in strongly driven systems. SFI, and TI in particular, is a process that is sensitive to the instantaneous field strength, making SFI a promising tool for time-resolving the study of ultrafast dynamics in molecules.
Furthermore, TI is the first step 
in the process of high-harmonic generation \cite{hhg1, hhg2}, a key mechanism underlying attosecond science.

There have been a number of efforts to develop approximate quantum mechanical theories of TI over the years, including the Ammosov-Delone-Krainov (ADK) model \cite{adk}, the extension of ADK for molecular cases (MOADK) \cite{moadk, moadk-kjeldsen-2005}, the Keldysh-Faisal-Reiss (KFR) method \cite{SFI_keldysh, keldysh-0000, faisal-1973, reiss-1980}, and the weak-field asymptotic theory (WFAT) \cite{wfat1, wfat2, wfat3, wfat4, wfat_diatomic_collection}
Alternatively, \rlremove{in direct numerical simulation} \rladd{in time-dependent wave function propagation methods \cite{tdcis, mctdhf_1, mctdhf_2, tdcasscf, tdrasscf, tdormas},} TI can be evaluated using
\rladd{absorbing masks \cite{hhg2}},
complex absorbing \rlremove{boundary} \rladd{potentials} \cite{cap-1986, cap-application-1989, cap-reflect-transmit-1996, cap-sfi-tdci-2014, cap-sfi-polyenes-2014, schlegel_o2_tdci-2015, cap-reduce-cost-2024},
\rladd{exterior complex scaling (ECS) \cite{ecs-he-2007, ecs-tao-2009}},
\rladd{infinite-range ECS \cite{irecs-2010, irecs-tdcasscf-2018}, or by projecting to scattering states 
\cite{extract-continuum-2007, pmd-circular-2016, interplay-pulse-param-2017, ocs-sfi-2020, ocs-ac-stark-2023}}
\rlremove{in time-dependent wave function propagation methods}. Among these theoretical methods, 
time-dependent evolution of either wave functions or densities \cite{tddft, tuned_rsx} are considered to be the most accurate. These methods, however, are often 
very expensive since at each time-step, one needs to construct the Hamiltonian from the two-electron integrals, which may reach more than several million for even medium-sized molecules and moderate-sized basis sets.
The alternatives to time-dependent methods are much cheaper at the expense of somewhat reduced accuracy. Of particular interest here, is the WFAT, which has been shown to give satisfactory agreement with experimental orientation-dependent ionization yields \cite{wfat-application-laser-elstru-2015, wfat-application-atto-cm-2015, endo2019angle, wfat-application-hhs-2024}. 

Several flavors of WFAT have been developed, as detailed in the next section. Briefly, the one-electron (OE-)WFAT \cite{wfat1,wfat2,wfat4,ir_oewfat} approximates the ionizing wave function with a single configuration framework while multiconfiguration effects can be treated in the many-electron (ME-)WFAT \cite{tr_mewfat,tr_mewfat_app,tr_mewfat_app2}. 
In WFAT, the ionization rate can be calculated using either the asymptotic portion of the wave function, called the tail representation (TR) \cite{wfat1, wfat2, tr_mewfat, tr_mewfat_app, tr_mewfat_app2}, or reformulated in the so-called integral representation (IR) \cite{ir_oewfat,ir_oewfat_hf,ir_oewfat_grid}.
In this work, we develop a
formulation of ME-WFAT 
that works in virtually any molecular geometry
by reformulating the theory in the integral representation (IR). 
Specifically, by reformulating the Schr\"{o}dinger equation in its integral form 
via the appropriate Green’s function, the resulting integral formulas allow Gaussian-type orbitals (GTO) to be used to expand the molecular orbitals (MO) without a significant loss
of accuracy. This greatly reduces the requirement on the accuracy of the asymptotic tail, and is the reason for the more general applicability of the IR formulation. Thus far, the IR has only been applied in the OE-WFAT formulation 
\cite{ir_oewfat}, where angle-dependent TI from molecules as big as benzene, naphtalene \cite{ir_oewfat_grid}, and phenyl rings system \cite{ir_oewfat_torsional} have been successfully calculated. IR ME-WFAT therefore paves the way towards an efficient computational method of TI that is applicable regardless of molecular geometry 
and is also capable of treating the effects due to multielectron interactions. 
The IR ME-WFAT method presented in this work is the foundation of the DFT-based ME-WFAT in Ref. \cite{wahyutama_irmewfat_dft},
where we showed that the resulting angle-dependent ionization yields are in excellent agreement with the full real-time time-dependent density functional theory calculations.

%
We illustrate IR-MEWFAT in a range of atoms, and small- to mid-size molecular systems. We show that the main way ME-WFAT improves angle-dependent ionization rates calculated by OE-WFAT is through the definition of the dipole moment used in the formulation and the use of Dyson orbital as the ionizing orbital. In particular, we show that angle-dependent ionization rates that are in good agreement with results from experiments or time-dependent methods in CO and N$_2$ can be recovered by using ME-WFAT. To demonstrate the versatility of the integral formulation for arbitrary molecular geometry, we demonstrate the computation of angle-dependent ionization rates in several polyatomic molecules, such as formic acid and CH$_3$F.
The computer codes used for running all simulations in this work, both ME-WFAT and OE-WFAT, are currently implemented in a
developer version of the \textsc{NWChem} quantum chemistry program \cite{nwchem, nwchem2}. 
\rladd{Another program that provides OE-WFAT capabilities is \textsc{PyStructureFactor} \cite{pystructurefactor-2023}, while the \textsc{SLIMP} package is developed to target more general properties of strong-field interactions with its ionization functionality implemented at the MOADK and KFR levels of theory \cite{slimp1-2015, slimp2-2024}.}

The rest of the paper is organized as follows: In section~\ref{sec:WFAT_background}, we detail the background in the development and comparison for the various flavors of WFAT. In section~\ref{sec:theory}
the formulation of ME-WFAT within IR will be outlined. We will present some illustrative calculations of structure factors in Section \ref{sec:results}, highlighting the capabilities of ME-WFAT relative to OE-WFAT. In Section \ref{sec:atoms}, we will discuss the equivalence between TR and IR ME-WFAT using some atoms as test cases. Section \ref{sec:o2_no} presents cases where OE-WFAT is sufficient for TI calculation. In Section \ref{sec:co_ocs} and \ref{sec:n2_hcooh_ch3f}, we discuss two main factors through which ME-WFAT improves OE-WFAT, these are the dipole moment and Dyson orbital. Finally, we will discuss prospective applications and improvements to IR ME-WFAT introduced in this work in Section \ref{sec:conclusions}.

\section{WFAT background \label{sec:WFAT_background}}


The earliest version of WFAT is the OE-WFAT, which is based on the single-active electron approximation \cite{wfat1, wfat2, wfat4}. Here, the defining problem starts from a one-electron, mean-field-type Schr\"odinger equation\rladd{, for example, the Roothan equation in Hartree-Fock (HF) or the Kohn-Sham equation in density functional theory (DFT)}. \rlremove{If using Hartree-Fock (HF), this is the Roothan equation, while if density functional theory (DFT) is used, this is the Kohn-Sham equation.}
Choosing the defining eigenproblem this way has two consequences: (i) 
Only one electron is involved in the rearrangement following the ionization that forms the final state
(the orbitals other than the ionizing one are frozen during ionization), and (ii) OE-WFAT can only be used in conjunction with single-determinant methods. This is because the orbital energy, needed in the calculation of ionization probability, is only well-defined for single-configuration wave functions. 
As a consequence of 
this, OE-WFAT can not simulate, even approximately, ionization starting from an excited initial state. To see why, recall that in a single-configuration framework, an excited state would be a configuration where an electron is moved from an occupied orbital to a virtual one. If, for example, the final state is the ground state (in which all electrons reside in the occupied orbitals), then the ionizing orbital is 
a virtual orbital of the initial charge state. Since virtual orbitals often have positive orbital energy, this violates the underlying assumption in WFAT that the initial energy should be negative.

WFAT has also been extended to formally include all electrons in the system, resulting in ME-WFAT \cite{tr_mewfat}. In contrast to OE-WFAT, ME-WFAT takes the exact all-electron Hamiltonian as the defining eigenproblem \cite{tr_mewfat}. This means the ionizing wave function is the fully antisymmetric, all-electron eigenfunction of the system. The problems with OE-WFAT mentioned in the preceding paragraph are absent
in ME-WFAT. Since the wave function is no longer restricted to a single-determinant, multiconfiguration effects in TI can now be treated \cite{tr_mewfat_app}.
For example, in high-harmonic generation, a process that involves TI in its initial step, it has 
been found that an adequate description of correlation is required to reproduce 
experiments that involve resonant enhancement via ionic states \cite{tdcasscf_rhhg, mo_rhhg}. 

Instead of orbital energy, ME-WFAT uses the proper ionization potential, that is, the difference between the initial and final state eigenvalues, which is always negative if the initial state lies below the final cationic state. Therefore, ionization from an excited initial state can be handled, in addition to ionization from the ground state. 

The original ME-WFAT introduced in Ref. \cite{tr_mewfat} is derived under TR where the orbitals need to have an accurate representation of the asymptotic dependence at large distance (the tail of the orbital). 
Because of this, the original formulation of ME-WFAT poses a numerical challenge---an accurate asymptotic dependence is difficult to achieve without using grid-type basis functions. A Hartee-Fock program based on  a specially designed grid exists \cite{hf_grid} but it is only formulated for atoms and diatomic molecules. In fact, designing a grid basis that is applicable to any polyatomic molecule 
while still incurring a tractable amount of computations when used in conjunction with accurate multiconfiguration methods is a great challenge. That is why it has been difficult to use TR ME-WFAT for any molecules other than the two geometry groups previously mentioned. Table \ref{tab:wfat_compare} summarizes the properties of the various representations of WFAT and whether one-electron or all-electron is used.

\begin{table*}[htb]
   \caption{\label{tab:wfat_compare} The comparison of the properties of WFAT types classified based on representation and the number of active electrons. In a hand-waving manner, going down the rows means broader range of molecular geometries is covered, while going to the right across columns means more accuracy because correlation becomes better described. Some related references for each type are also given. 
   }
   \begin{tabular}{!{\vrule width 1.2pt} c !{\vrule width 1.2pt} c|c !{\vrule width 1.2pt}}
     \hlineadj{1.2pt}
     & & \\[-0.7em]
     & OE-WFAT & ME-WFAT \\
     & & \\[-0.7em]
     \hlineadj{1.2pt}
     & & \\
     \hspace{2mm} TR \hspace{2mm} 
     & 
     \hspace{2mm}
     \begin{tabular}{@{}l@{}}
     - Only works with single-configuration method \\
     - Can (approximately) handle excited final state \\
     - Cannot handle excited initial state \\
     - No $2e$ integrals \\ 
     - Works best for atoms and diatomic molecules \\
     Ref. \cite{wfat1, wfat2, wfat3, wfat4}.
     \end{tabular}  
     \hspace{2mm}
     & 
     \hspace{2mm}
     \begin{tabular}{@{}l@{}}
     - Works with any number of configurations \\
     - Can handle excited initial and final states \\
     - No $2e$ integrals \\ 
     - Works best for atoms and diatomic molecules \\
     Ref. \cite{tr_mewfat, tr_mewfat_first_order, tr_mewfat_app}.
     \end{tabular} 
     \hspace{2mm} \\
     & & \\[-0.5em]
     \hline
     & & \\[-0.5em]
     \hspace{2mm} IR \hspace{2mm} 
     & 
     \begin{tabular}{@{}l@{}}
     - Only works with single-configuration method \\ 
     - Can (approximately) handle excited final state \\
     - Cannot handle excited initial state \\
     - Need to evaluate $2e$ integrals \\
     - No restriction on the molecular geometry \\
     Ref. \cite{ir_oewfat, ir_oewfat_hf, ir_oewfat_grid}.
     \end{tabular} 
     \hspace{2mm}
     & 
     \hspace{2mm}
     \begin{tabular}{@{}l@{}}
     - Works with any number of configurations \\ 
     - Can handle excited initial and final states \\
     - Need to evaluate $2e$ integrals \\
     - No restriction on the molecular geometry \\
     \textit{This work}.
     \end{tabular} 
     \hspace{2mm} \\
      & & \\
     \hlineadj{1.2pt}
   \end{tabular}
\end{table*}


\section{Theory} \label{sec:theory}

\subsection{The \textit{first} asymptotic expansion of multielectron wave functions}\label{sec:asymptotic_expansion}


We consider a many-electron molecule perturbed by an external electric field whose magnitude is such that 
single ionization is the dominant process
We then aim to establish a set of formulas that allow us to calculate the ionization rate 
as a function of field strength and the relative orientation between the molecule and the field. For one-electron WFAT, the starting equation is the Schr\"odinger equation for an electron subjected to an effective one-body potential, and this is often the mean-field equation that defines the orbitals. In the case of many-electron WFAT, we start from the original Schr\"odinger equation of an $N$-electron system,
\begin{align}
    \label{eq:schroedinger}
    H^{(N)} \Psi_n(Q_N) = E_n^{(N)} \Psi_n(Q_N),
\end{align}
where
\begin{align}
    H^{(N)} =& \, 
    -\frac{1}{2} \sum_{i=1}^N \nabla_i^2 + V^{(N)}, \\
    V^{(N)}(\mathbf R_N) =& -
    \sum_{i=1}^N \sum_{I=1}^{N_A} \frac{Z_I}{|\mathbf r_i - \mathbf C_I|} +
    \sum_{i=1}^{N-1} \sum_{j=i+1}^N \frac{1}{|\mathbf r_i - \mathbf r_j|} \nonumber\\
    &
    +\sum_{i=1}^N Fz_i ,
\end{align}
where $F$ is the magnitude of the external static field pointing in the positive $z$ direction, $N_A$ is the number of nuclei, $\mathbf C_1, \, \mathbf C_2, \ldots, \mathbf C_{N_A}$ are the nuclear coordinates, and $Z_1, \, Z_2, \, \ldots, Z_{N_A}$ are their charges. The following notations for the coordinates are used throughout this paper,
\begin{subequations}
   \begin{align}
       Q_N &\equiv \{ q_1, \ldots, q_{N-1},    q\}, \\
       \mathbf R_N &\equiv \{ \mathbf r_1, \ldots, \mathbf r_{N-1},    \mathbf r\}, \\
       q_i &\equiv \{\mathbf r_i, s_i\}, \\
       q &\equiv q_N.   
   \end{align}
\end{subequations}
where $\mathbf r_i \equiv \{r_i,\theta_i,\varphi_i\}$ and $s_i$ are the spatial and spin coordinates of the $i$-th electron. In the rest of this paper, $q$ is synonymous to $q_N$ (the coordinate of the last electron). It also proves to be advantageous to work in parabolic coordinate for the last electron since it allows for the separation of Schr\"odinger equation with a static field in the asymptotic region,
\begin{subequations}
   \begin{align}
       \xi  =& \,\, r + z, \hspace{1cm} 0\leq  \xi < \infty\\
       \eta =& \,\, r - z, \hspace{1cm} 0\leq \eta < \infty\\
       \varphi =& \,\, \operatorname{arctan}\frac{y}{x}, \hspace{1cm} 0\leq \varphi \leq 2\pi.
   \end{align}
\end{subequations}
In what follows, we chose the last electron to be the one removed due to single ionization. With the static external force pulling the electrons in the negative $z$ direction, 
the ionization region (which lies in the asymptotic region) is therefore represented by $\eta\to\infty$. We will also conventionally refer to the initial molecule as the `neutral'  and the ionized one as the \lq cation\rq. However, the derivation below applies to any initial and final charge states as long as the latter has one less electron than the former.

Since we are considering single ionization, it is useful to use an \textit{ansatz} in which the eigenfunctions of Eq. \eqref{eq:schroedinger} 
are expanded in terms of a basis that factorizes into an antisymmetrized product between an $(N-1)$-electron function and a one-electron function,
\begin{align}
    \label{eq:expand_antisym}
    \Psi_n(Q_N) =
    \sum_{n'} 
    \frac{N!}{(N-1)!} \,
    \hat{\mathcal{A}}\left(\tilde\Psi_{n'}(Q_{N-1}) t_{n'n}(q)\right),
\end{align}
where $\hat{\mathcal{A}}$ is the antisymmetrization operator. 
The function $t_{n'n}(q)$, in which all information about the ionized electron will be encoded, is expressed as a linear combination of \textit{parabolic channel functions} $\Phi_\nu^{n'n}(\xi,\varphi)$, 
\begin{align}
     \label{eq:t_expand}
     t_{n'n}(\eta,\xi,\varphi,s) = 
     \frac{1}{\sqrt{\eta}}
     \sum_{\nu}
     F_{\nu\sigma}^{n'n}(\eta) 
     \Phi_\nu^{n'n}(\xi,\varphi) \chi_\sigma(s),
\end{align}
and is the unknown that we will partially solve in the following analysis (it will be a partial solution because we are only interested in the asymptotic region relevant for ionization). The sum in Eq. \eqref{eq:t_expand} runs over all parabolic quantum number pairs $\nu\equiv(n_\xi,m)$, $n_\xi=0,1,\ldots$, $m=0,\pm 1, \ldots$ \cite{wfat1, wfat2}, and $\sigma\in\{\alpha,\beta\}$ is the $z$-projection of the spin angular momentum, hence $\alpha=1/2$ and $\beta=-1/2$. 
$\sigma$ implicitly depends on $n'$ due to the requirement that, together with the spin projection of $\tilde\Psi_{n'}$, it must produce the spin projection of $\Psi_n$.
Eq. \eqref{eq:t_expand} is the many-electron generalization of Eq. (11) in Ref. \cite{wfat3}. The parabolic channel function is the solution of the following eigen equation
\begin{align}
    \label{eq:adiabatic_equation}
    \mathcal{B}^{n'n} \Phi_\nu^{n'n}(\xi,\varphi) = \beta_\nu^{n'n} \Phi_\nu^{n'n}(\xi,\varphi),
\end{align}
where
\begin{align*}
    \mathcal{B}^{n'n} =& \, 
    \frac{\partial}{\partial \xi} \left( \xi \frac{\partial}{\partial \xi} \right) + 
    \frac{1}{4\xi} \frac{\partial^2}{\partial \varphi^2} +
    \frac{\textrm{IP}_{n'n}\xi}{2} - 
    \frac{F\xi^2}{4} + 
    Z_c ,
\end{align*}
and thus may be written as
\begin{align}
    \label{eq:adiabatic_define}
    \Phi_\nu^{n'n}(\xi,\varphi) = \phi_{n_\xi|m|}^{n'n} (\xi)
    \frac{e^{im\varphi}}{\sqrt{2\pi}}.
\end{align}

Several remarks on the form of Eq. \eqref{eq:expand_antisym} are in order. This equation is in a general form because, even though it has been expressed in the most suitable way for single ionization, it still allows for a straightforward generalization to higher order ionization. If simultaneous double ionization is desired, for instance, one can reexpress Eq. \eqref{eq:expand_antisym} so that the basis is in the form of $\hat{\mathcal A} ( \tilde\Psi_{n'}(Q_{N-2}) t_{n'n}(q, q') )$.
The expansion in Eq. \eqref{eq:expand_antisym} also includes the localized and delocalized contributions to $\Psi_n(Q_N)$. An example of localized contributions would be the contributions from the bound states of the unperturbed neutral molecule. In this case, the densities of both $\tilde\Psi_{n'}$ and $t_{n'n}$ are localized near the nuclei. The case of ionization focuses on the delocalized terms on the RHS of Eq. \eqref{eq:expand_antisym}. 

Evaluating Eq. \eqref{eq:expand_antisym} in the asymptotic region ($\eta \rightarrow \infty$) will quench 
any terms on the RHS for which $t_{n'n}(q)$ is localized, leaving 
only terms for which $t_{n'n}(q)$ is delocalized,  while $\tilde\Psi_{n'}(Q_{N-1})$ can be either
localized or delocalized.  Imposing the single ionization approximation will suppress the latter, hence we are left with terms for which $t_{n'n}(q)$ is delocalized and $\tilde\Psi_{n'}(Q_{N-1})$ is localized. Since $\tilde\Psi_{n'}(Q_{N-1})$ is localized, without loss of generality, we can use the eigenstates of the $(N-1)$-electron Hamiltonian in its place, i.e. $\tilde\Psi_{n'} \rightarrow \Psi_{n'}^+$, where $\Psi_{n'}^+$ is an eigenstate of $H^{(N-1)}$. Because $\Psi_{n'}^+$ is local, in the region $\eta\to\infty$ the exchange interaction between the ionized electron and the bound ones dies
away, and the antisymmetric product in Eq. \eqref{eq:expand_antisym} reduces to an ordinary product,
\begin{align}
    \label{eq:antisym_asymptotic}
    \hat{\mathcal{A}}\big( \Psi_{n'}^+(Q_{N-1}) 
    &t_{n'n}(q) \big) \big|_{\eta \to \infty}
    \approx  \nonumber \\
    &
    \frac{(N-1)!}{N!} 
    \Psi_{n'}^+(Q_{N-1}) \tau_{n'n}(q),
\end{align}
where we have defined $\tau_{n'n}(q) \equiv t_{n'n}(q)\big|_{\eta\to\infty}$. Therefore, in the asymptotic region Eq. \eqref{eq:expand_antisym} has the form of
\begin{align}
    \label{eq:expand_asymptotic}
    \Psi_n(Q_N)\big|_{\eta\to\infty} = 
    \sum_{n'} 
    \Psi_{n'}^+(Q_{N-1}) \tau_{n'n}(q) .
\end{align}
Note that on the RHS, only $\tau_{n'n}(q)$ is evaluated asymptotically, whereas the final eigenstate $\Psi_{n'}^+$ stays bound.

In order to solve for $\tau_{n'n}(q)$, the first step is to substitute the asymptotic wave function in Eq. \eqref{eq:expand_asymptotic} into Eq. \eqref{eq:schroedinger} in the asymptotic region,
\begin{align}
    \label{eq:schoredinger_asymptotic}
    H^{(N)}\big|_{\eta \to \infty} \Psi_n(Q_N)\big|_{\eta\to\infty} =
    E_n^{(N)} \Psi_n(Q_N) \big|_{\eta\to\infty}.
\end{align}
where
\begin{gather}
    \label{eq:asymptotic_hamiltonian}
    H^{(N)}\big|_{\eta \to \infty} =
    H^{(N-1)} - 
    \frac{\nabla^2}{2} - 
    \frac{Z_c}{r} +
    O(r^{-2}) + Fz  ,\\
    Z_c = \sum_{I=1}^{N_A} Z_I - N + 1, \nonumber
    %
\end{gather}
We note that the terms represented by $O(r^{-2})$ in Eq. \eqref{eq:asymptotic_hamiltonian} above contains the dipole term of the asymptotic expansion of $V^{(N)}(\mathbf R_N)$. This dipole term will be omitted from this point onward because we are interested in the lowest order approximation of ME-WFAT. This term becomes important when one considers the first order correction \cite{tr_mewfat_first_order}.
By using Eq. \eqref{eq:expand_asymptotic} in Eq. \eqref{eq:schoredinger_asymptotic} and then projecting to a particular final eigenstate, one obtains
\begin{gather}
    \label{eq:t_equation}
    \left(
    -
    \frac{\nabla^2}{2} 
    - 
    2\,\frac{Z_c}{\eta+\xi} 
    -
    \frac{1}{2}F(\eta-\xi)  
    -
    \textrm{IP}_{n'n}
    \right) 
    \tau_{n'n}(q)
    = 0
\end{gather}
where $\textrm{IP}_{n'n} = E_n^{(N)} - E_{n'}^{(N-1)}$ is the ionization potential, and we have applied the change of coordinate to the parabolic coordinate system.

Using Eq. \eqref{eq:t_expand}, \eqref{eq:adiabatic_equation}, and \eqref{eq:adiabatic_define}, Eq. \eqref{eq:t_equation} can be simplified to an equation satisfied by $F_{\nu\sigma}^{n'n}(\eta)\big|_{\eta\to\infty}$,
\begin{align}
    \label{eq:f_equation}
    \left(
    \frac{\partial^2}{\partial\eta^2} +
    \frac{F\eta}{4} +
    \frac{\textrm{IP}_{n'n}}{2}
    \right) F_{\nu\sigma}^{n'n}(\eta)\big|_{\eta\to\infty} = 0
\end{align}
where the $O(\eta^i)$ terms with $i=-1,-2$ have been dropped since they are negligible in the asymptotic region. Eq. \eqref{eq:f_equation} can be transformed to the Airy differential equation and thus its solution can be expressed in terms of the asymptotic form of the Airy function,
\begin{align}
    F_{\nu\sigma}^{n'n}(\eta)\big|_{\eta\to\infty} =
    & \,\,
    \frac{\sqrt{2} \, f_{\nu\sigma}^{n'n}}{(F\eta)^{1/4}}
    \exp\left(
    \frac{i\sqrt{F}}{3}\eta^{3/2} + 
    \frac{i\textrm{IP}_{n'n}}{\sqrt{F}}\sqrt{\eta}
    \right).
    \label{eq:f_solution}
\end{align}
Eq. \eqref{eq:expand_asymptotic} along with Eq. \eqref{eq:t_expand} and \eqref{eq:f_solution} constitute the asymptotic expansion of the perturbed neutral wave function $\Psi_n(Q_N)$ interacting with an external static field pointing in the positive $z$ direction in terms of some asymptotic basis functions weighted by $f_{\nu\sigma}^{n'n}$. Since in WFAT this coefficient is interpreted as the ionization probability, the goal of any WFAT derivation is then to construct an expression for the expansion coefficients $f_{\nu\sigma}^{n'n}$ in terms of the field strength and orientation, as well as the unperturbed properties of the system such as the unperturbed ionization energy, dipole moment, and wave functions. The next section is devoted to obtaining such an expression for $f_{\nu\sigma}^{n'n}$ within the integral formulation of ME-WFAT.

\subsection{The \textit{second} asymptotic expansion of multielectron wave functions}
In Section \ref{sec:asymptotic_expansion}, we 
derived 
an asymptotic expansion of the initial wave function interacting with a static electric field pointing in the positive $z$ direction (also shown in Eq. \eqref{eq:wfn_expand1} below).  In this section, we will first obtain another form of the asymptotic expansion for $\Psi_n(Q_N) \big|_{\eta\to\infty}$ different from that derived in Section \ref{sec:asymptotic_expansion}. But since these two expansions are equivalent, comparing them should yield the sought expression for $f_{\nu\sigma}^{n'n}$.

As the first step, note that Eq. \eqref{eq:schroedinger} can be rewritten
as
\begin{align}
    \label{eq:schroedinger_ref}
    \left(H_{\textrm{ref}} - E_n^{(N)}\right) \Psi_n(Q_N) = -V_c(\mathbf R_N) \Psi_n(Q_N)
\end{align}
where
\begin{gather}
    H_{\textrm{ref}} 
    = \,H^{(N-1)} - 
    \frac{\nabla^2}{2} - 
    \frac{Z_c}{r} +
    Fz + 
    O(r^{-2}), 
    \label{eq:href_define}\\
    V_c(\mathbf R_N) 
    = \, V_{1e}(\mathbf r) + V_{2e}(\mathbf R_N) +
    O(r^{-2}),
    \label{eq:vc_v1e_v2e} \\
    V_{2e}(\mathbf R_N) 
    = \sum_{i=1}^{N-1} \frac{1}{|\mathbf r - \mathbf r_i|},  \label{eq:v2e_define} \\
    \label{eq:v1e_define}
    V_{1e}(\mathbf r) 
    = -\sum_{I=1}^{N_A} \frac{Z_I}{|\mathbf r - \mathbf C_I|}
    +
    \frac{Z_c}{r}.
\end{gather}
We note that the $O(r^{-2})$ terms in Eq. \eqref{eq:href_define} and \eqref{eq:vc_v1e_v2e} are identical, and are incorporated to account for the dipole term of $V^{(N)}(\mathbf R_N)$. For the same argument regarding the $O(r^{-2})$ term in Eq. \eqref{eq:asymptotic_hamiltonian}, this higher multipole term will be omitted in this work.
One might notice that the various potential terms in Eq. \eqref{eq:vc_v1e_v2e}-\eqref{eq:v1e_define} are not entirely origin-independent. This is because of the multipole expansion of the core potential, which gives
rise to the origin-dependent terms, such as the $Z_c/r$ and the dipole potential term contained in $O(r^{-2})$. As has been shown in Ref. \cite{wfat2, ir_oewfat_hf}, for wave functions having the exact asymptotic behavior, the structure factor and hence the ionization rate are origin-independent (see also Section \ref{sec:asymptote_cancel}).

In the integral form, Eq. \eqref{eq:schroedinger_ref} (or equivalently, Eq. \eqref{eq:schroedinger}) takes the form 
\begin{align}
    \label{eq:integral_net_wfn}
    \Psi_n(Q_N) = 
    -\int dQ'_N \,
    G_n(Q_N' ; Q_N) V_c(\mathbf R'_N) \Psi_n(Q'_N)
\end{align}
where the Green's function $G_n(Q_N' ; Q_N)$ is the solution of the following inhomogeneous equation
\begin{align}
    \label{eq:green_equation}
    \left(H_{\textrm{ref}} - E_n^{(N)}\right)  G_n(Q_N' ; Q_N) = \prod_{i=1}^N \delta(\mathbf r'_i - \mathbf r_i) \, \delta_{s'_i s_i}.
\end{align}
We note that the integral solution of the type of Eq. \eqref{eq:integral_net_wfn} is the corner stone of the integral representation of OE-WFAT and ME-WFAT introduced in the present work. Because this integral integrates over the entire $4N$ dimensions, one is not presented with the difficulty found in the TR associated with the incomplete integration leaving the $\eta$ coordinate unintegrated. This leads to the TR ionization rates becoming sensitive to the accuracy of the asymptotic tail (that occupies the unintegrated dimension) of the wave function.

The next step is to solve for $G_n(Q'_N ; Q_N)$ for which we will use the following spectrally resolved \textit{ansatz},
\begin{align}
    \label{eq:green_expand}
    G_n(Q'_N ; Q_N) =
    & \,
    -\frac{2}{\sqrt{\eta' \eta}}
    \sum_{n'\nu} 
    \Psi_{n'}^+(Q_{N-1}) 
    \left(\Psi_{n'}^+(Q'_{N-1})\right)^* \nonumber
    \\
    & \,
    \mathcal{G}_{\nu}^{n'n}(\eta,\eta')
    \Phi_\nu^{n'n}(\xi,\varphi)   
    \Phi_{\Bar{\nu}}^{n'n}(\xi',\varphi')       
    \chi_\sigma(s) \nonumber
    \\
    & \,
    \chi_\sigma^\dagger(s')
\end{align}
where $\Bar{\nu} \equiv (n_\xi,-m)$ and $\mathcal G_{\nu}^{n'n}(\eta,\eta')$ is a one-dimensional Green's function that we are going to solve for. 
Substituting Eq. \eqref{eq:green_expand} into Eq. \eqref{eq:green_equation}, and following similar steps to Ref. \cite{ir_oewfat}, one arrives at the following differential equation
\begin{gather}
    \left( 
    \frac{\partial^2}{\partial\eta^2} 
    + 
    \frac{1-m^2}{4\eta^2} 
    +
    \frac{\textrm{IP}_{n'n}}{2} 
    +
    \frac{F\eta}{4}
    + 
    \frac{\beta_\nu^{n'n}}{\eta}
    \right) 
    \mathcal G_\nu^{n'n}(\eta,\eta') \nonumber\\
    = \delta(\eta-\eta').
    \label{eq:alpha_equation}
\end{gather}
As a side note, the differential operator in the LHS of Eq. \eqref{eq:alpha_equation} is identical to the one that gives rise to Eq. \eqref{eq:f_equation} before omitting the $O(\eta^i)$ terms, with $i=-1,-2$.

In order so that $\mathcal G_\nu^{n'n}(\eta,\eta')$ is uniquely defined through Eq. \eqref{eq:alpha_equation}, we must impose boundary conditions on this equation. A physically meaningful choice is that $\mathcal G_\nu^{n'n}(\eta,\eta') \big|_{\eta\to 0} \in [0,\infty)$ and $\mathcal G_\nu^{n'n}(\eta,\eta') \big|_{\eta\to\infty} = 0$. We also denote the two linearly independent solutions to the homogeneous equation that arises from Eq. \eqref{eq:alpha_equation} when $\eta\neq\eta'$ as $R_\nu^{n'n}(\eta)$ and $O_\nu^{n'n}(\eta)$. We then arbitrarily assign $R_\nu^{n'n}(\eta)$ to the solution that is regular at $\eta=0$, hence $0 \leq R_\nu^{n'n}(0) < \infty$, and $O_\nu^{n'n}(\eta)$ to the one that vanishes as $\eta\to\infty$, hence $\lim_{\eta\to\infty} O_\nu^{n'n}(\eta) = 0$. 
There are a number of known methods for solving the Green's function of a second order differential equation \cite{arfken_weber}. Using these methods,
it can be shown that the solution of Eq. \eqref{eq:alpha_equation}, satisfying the aforementioned boundary conditions, is
\begin{align}
    \label{eq:alpha_general}
    \mathcal G_\nu^{n'n}(\eta,\eta') =
    \left\{
    \begin{array}{ll}
        O_\nu^{n'n}(\eta') R_\nu^{n'n}(\eta) / \mathcal W_\nu^{n'n} &,   0 \leq \eta < \eta'\\
        & \\
        R_\nu^{n'n}(\eta') O_\nu^{n'n}(\eta) / \mathcal W_\nu^{n'n} &,   \eta > \eta'
    \end{array}
    \right.
\end{align}
with the Wronskian $\mathcal W_\nu^{n'n}$ defined to be
\begin{align}
    \mathcal W_\nu^{n'n} =&   \,\,
    R_\nu^{n'n}(\eta) \frac{d O_\nu^{n'n}(\eta)}{d\eta} - 
    O_\nu^{n'n}(\eta) \frac{d R_\nu^{n'n}(\eta)}{d\eta} \nonumber \\
    =& \,\, \textrm{constant}
\end{align}
The Wronskian being constant follows from the absence of the first order derivative term in the differential Eq. \eqref{eq:alpha_equation} \cite{arfken_weber}. From Eq. \eqref{eq:alpha_general}, we obtain
\begin{align}
    \label{eq:alpha_asymptotic0}
    \mathcal G_\nu^{n'n}(\eta,\eta') \Big|_{\eta\to\infty} = 
    \frac{R_\nu^{n'n}(\eta')}{\mathcal W_\nu^{n'n}} O_\nu^{n'n}(\eta) \Big|_{\eta\to\infty}.
\end{align}

To obtain $O_\nu^{n'n}(\eta) \big|_{\eta\to\infty}$, we note that it is a solution of Eq. \eqref{eq:alpha_equation} when the RHS is set to zero and when $\eta\to\infty$. But these conditions make this equation identical to Eq. \eqref{eq:f_equation}, implying that $O_\nu^{n'n}(\eta) \big|_{\eta\to\infty}$ must have the same form as Eq. \eqref{eq:f_solution} with the only difference being in the constant prefactor (given that Eq. \eqref{eq:f_equation} is homogeneous). This means, in Eq. \eqref{eq:f_solution}, we can arbitrarily make a replacement $f_{\nu\sigma}^{n'n} \to O_\nu^{n'n}$ in the RHS, and identify the LHS as $O_\nu^{n'n}(\eta) \big|_{\eta\to\infty}$. Applying this to Eq. \eqref{eq:alpha_asymptotic0} results in

\begin{align}
    \mathcal G_\nu^{n'n}(\eta,\eta') &\Big|_{\eta\to\infty} = \,\, 
    \frac{O_\nu^{n'n}}{\mathcal W_\nu^{n'n}}
    R_\nu^{n'n}(\eta')
    \frac{\sqrt{2}}{(F\eta)^{1/4}}  \nonumber \\
    &\,\,
    \times
    \exp\left(
    i\frac{\sqrt{F}}{3}\eta^{3/2} +
    i\frac{\textrm{IP}_{n'n}}{\sqrt{F}} \eta^{1/2}
    \right).
    \label{eq:alpha_asymptotic1}
\end{align}
By substituting Eq. \eqref{eq:alpha_asymptotic1} into \eqref{eq:green_expand}, and then Eq. \eqref{eq:green_expand} into \eqref{eq:integral_net_wfn}, we have essentially arrived at a new asymptotic expansion for $\Psi_n(Q_N)$ with the unknown being the coefficient $O_\nu^{n'n}$. To facilitate a clearer picture of where we are within the derivation process so far, we will explicitly write down these two different-looking but equivalent asymptotic expansions. The first one, which is formed with the help of Eq. \eqref{eq:expand_asymptotic}, \eqref{eq:t_expand}, and \eqref{eq:f_solution}, reads
\begin{widetext}
   \begin{align}
       \label{eq:wfn_expand1}
       \Psi_n(Q_N) \big|_{\eta\to\infty} =& \,\,
       \frac{1}{\sqrt{\eta}}
       \sum_{n'}
       \frac{\sqrt{2}}{(F\eta)^{1/4}}
       \exp\left(
       i\frac{\sqrt{F}}{3}\eta^{3/2} + 
       i\frac{\textrm{IP}_{n'n}}{\sqrt{F}}\eta^{1/2}
       \right)
       \Psi_{n'}^+(Q_{N-1})
       \sum_{\nu\sigma}
       \Phi_\nu^{n'n}(\xi,\varphi) \,
       \chi_\sigma(s)
       f_{\nu\sigma}^{n'n},
   \end{align}
   while the second one, which is formed through the use of Eq. \eqref{eq:integral_net_wfn}, \eqref{eq:green_expand}, and \eqref{eq:alpha_asymptotic1}, reads
   \begin{align}
       \label{eq:wfn_expand2}
       \Psi_n(Q_N) \big|_{\eta\to\infty} =& \,\,
       \frac{1}{\sqrt{\eta}}
       \sum_{n'}
       \frac{\sqrt{2}}{(F\eta)^{1/4}}
       \exp\left(
       i\frac{\sqrt{F}}{3}\eta^{3/2} + 
       i\frac{\textrm{IP}_{n'n}}{\sqrt{F}}\eta^{1/2}
       \right)
       \Psi_{n'}^+(Q_{N-1})
       \sum_{\nu\sigma}
       \Phi_\nu^{n'n}(\xi,\varphi) 
       \chi_\sigma(s)   \nonumber\\
       & \,\,
       \times
       \left\{
       \frac{2O_\nu^{n'n}}{\mathcal W_\nu^{n'n}} 
       \int dQ'_N
       \frac{R_\nu^{n'n}(\eta')}{\sqrt{\eta'}} \Phi_{\Bar{\nu}}^{n'n}(\xi',\varphi')
       \chi_\sigma^\dagger(s')
       \left( \Psi_{n'}^+(Q'_{N-1}) \right)^*
       V_c(\mathbf R'_N)
       \Psi_n(Q'_N)
       \right\}.
   \end{align}
\end{widetext}
Eq. \eqref{eq:wfn_expand1} and \eqref{eq:wfn_expand2} have been arranged such that all common quantities are written first. Comparing these two equations, it is clear that the coefficient $f_{\nu\sigma}^{n'n}$ is equal to the terms inside the curly brackets in Eq. \eqref{eq:wfn_expand2}, which may also be understood to be the many-electron generalization of Eq. (54) in Ref. \cite{ir_oewfat}.

\subsection{The leading order weak-field approximation}
Despite having obtained a closed form expression for $f_{\nu\sigma}^{n'n}$, the various terms inside the curly brackets in Eq. \eqref{eq:wfn_expand2}, such as the initial and final wave functions, among others, are still not yet straightforwardly calculable because these terms are associated with a Hamiltonian that has an external DC field term. In addition, the corresponding Schr\"odinger equation for general multielectron systems cannot be easily solved 
due to the need of the continuum eigensubspace of the unperturbed Hamiltonian. This is the point where the weak-field approximation is invoked.

In this approximation, the field is assumed to be weak enough
compared to the binding force of the nuclei, therefore the quantities that depend on the field strength can be expanded in 
powers of the field, the same way any
perturbation-type expansion is performed, namely
\begin{subequations}
   \label{eq:perturb_expansion}
   \begin{align}
       \Psi_n &= \Psi_n^{(0)} + F \Psi_n^{(1)} + O(F^2) \\
       \Psi_n^+ &= \Psi_n^{+(0)} + F \Psi_n^{+(1)} + O(F^2) \\
       E_n^{(N)} &= E_n^{(N)(0)} - F\mu_{nz}^{(N)} + O(F^2) \\
       R_\nu^{n'n} &= R_\nu^{n'n(0)} + F R_\nu^{n'n(1)} + O(F^2) \\
       \Phi_\nu^{n'n} &= \Phi_\nu^{n'n(0)} + F \Phi_\nu^{n'n(1)} + O(F^2) \\
       \beta_\nu^{n'n} &= \beta_\nu^{n'n(0)} + F \beta_\nu^{n'n(1)} + O(F^2) 
   \end{align}
\end{subequations}
and a similar definition as $E_n^{(N)}$ applies to $E_{n'}^{(N-1)}$.
The 
Wronskian, within the weak-field approximation, can be shown to be 
\begin{align}
    \label{eq:wronskian}
    \mathcal W_\nu^{n'n} = -\varkappa = 
    -\sqrt{2\left|\textrm{IP}_{n'n}^{(0)}\right|}
\end{align}
with $\textrm{IP}_{n'n}^{(0)} = E_n^{(N)(0)} - E_{n'}^{(N-1)(0)} $,  the field-free ionization potential \cite{ir_oewfat}. 
$\mu_{nz}^{(N)}$ is the $z-$component of $\boldsymbol \mu_n^{(N)}$, the total electronic dipole moment corresponding to $\big|\Psi_n^{(0)} \big\rangle$ in the lab-frame. Apart from $\mu_{nz}^{(N)}$, we will also need $\boldsymbol \mu_{n'z}^{(N-1)}$, the total electronic dipole moment of $\big|\Psi_{n'}^{+(0)}\big\rangle$ in the lab-frame. These dipole vectors are given by
\begin{gather}
    \boldsymbol{\mu}_n^{(N)} = -\Big\langle \Psi_n^{(0)} \Big| \sum_{i=1}^N \hat{\mathbf r}_i \Big| \Psi_n^{(0)} \Big\rangle, \label{eq:neutral_emoment}
    \\
    \boldsymbol{\mu}_{n'}^{(N-1)} = -\Big\langle \Psi_{n'}^{+(0)} \Big| \sum_{i=1}^{N-1} \hat{\mathbf r}_i \Big| \Psi_{n'}^{+(0)} \Big\rangle. \label{eq:cation_emoment}
\end{gather}

The $O_\nu^{n'n}$ coefficient has been solved by employing the \textit{connection formula} in a previous work on WFAT \cite{wfat3}. The derivation in the case of IR ME-WFAT is completely parallel to that in the aforementioned reference, with the only difference being the definition of orbital energy and dipole moments involved in the derivation. We therefore refer the readers to these references for the details of obtaining $O_\nu^{n'n}$. Here, we simply invoke the final result for the leading order term, which reads
\begin{align}
    O_\nu^{n'n(0)} =& \,\,
    \sqrt{\frac{\varkappa}{2}}
    \left(
    \frac{4\varkappa^3}{F}
    \right) ^ {\beta_\nu^{n'n(0)}/\varkappa}  \nonumber \\
    &\,\,
    \times
    \exp\left(
    i\frac{\pi}{4} +
    i\frac{\pi\beta_\nu^{n'n(0)}}{\varkappa} -
    \mu_z^{n'n} \varkappa -
    \frac{\varkappa^3}{3F}
    \right),
    \label{eq:o_zeroth}
\end{align}
where $\beta_\nu^{n'n(0)} = Z_c - \varkappa \left( n_\xi + \frac{|m|+1}{2} \right)$
and $\mu_z^{n'n}$ is the lab-frame $z$-component of the vector \rladd{difference} $\boldsymbol{\mu}_n^{(N)} - \boldsymbol{\mu}_{n'}^{(N-1)}$. This appearance of \rlremove{the} dipole moment difference in ME-WFAT within the integral representation \rlremove{can also be found in} \rladd{is consistent with} the tail representation \cite{tr_mewfat}\rladd{, and can also be found in other TI theories that incorporate Stark shift correction \cite{sfi-polar-stark-2010, co-no-orient-two-color-2011, dipole-influence-2017}}. 
We note 
that the factor $(4\varkappa^3/F)^{\beta_\nu^{n'n(0)}/\varkappa}$ in Eq. \eqref{eq:o_zeroth} is different from that in Eq. (50) in Ref. \cite{ir_oewfat} in which the power of $\varkappa$ is $2$, instead of $3$. As can be checked by rederiving Eq. (59) in the aforementioned paper, the correct power of $\varkappa$ is $3$ (as appears in Eq. \eqref{eq:o_zeroth} above). 

Now that we have everything we need to construct the leading order approximation to the coefficient $f_{\nu\sigma}^{n'n}$, we can substitute Eq. \eqref{eq:wronskian}, Eq. \eqref{eq:o_zeroth}, and the zeroth order terms in Eq. \eqref{eq:perturb_expansion} for the corresponding terms inside the curly brackets in Eq. \eqref{eq:wfn_expand2}. This yields,
\begin{align}
    \label{eq:f_coeff}
    f_{\nu\sigma}^{n'n(0)} =& \,\, 
    \sqrt{\frac{\varkappa}{2}}
    \left(
    \frac{4\varkappa^2}{F}
    \right) ^ {\beta_\nu^{n'n(0)}/\varkappa} g_{\nu\sigma}^{n'n}(\beta,\gamma)  \nonumber \\
    &\,\,
    \times
    \exp\left(
    i\frac{\pi}{4} +
    i\frac{\pi\beta_\nu^{n'n(0)}}{\varkappa} -
    \mu_z^{n'n} \varkappa -
    \frac{\varkappa^3}{3F}
    \right)
\end{align}
where the \textit{asymptotic coefficient} $g_{\nu\sigma}^{n'n}$ is defined as
\begin{align}
    \label{eq:asymptotic_coeff0}
    g_{\nu\sigma}^{n'n}(\beta,\gamma) = 
    & \,
    -2 \varkappa^{\beta_\nu^{n'n(0)}/\varkappa - 1} 
    \int dQ_N \,\,
    \frac{R_\nu^{n'n(0)}(\eta)}{\sqrt{\eta}}    \nonumber
    \\
    &
    \times
    \Phi_{\Bar{\nu}}^{n'n(0)}(\xi,\varphi) \,
    \chi_{\sigma}^\dagger(s) 
    \left(\Psi_{n'}^{+(0)}(Q_{N-1})\right)^*    \nonumber\\
    &
    \times
    V_c(\mathbf R_N) \Psi_n^{(0)}(Q_N).
\end{align}
Of particular interest is the structure factor, defined by
\begin{align}
    G_{\nu\sigma}^{n'n} (\beta,\gamma) = e^{-\varkappa\mu_z^{n'n}} \, g_{\nu\sigma}^{n'n} (\beta,\gamma),
\end{align}
which is a field-independent quantity and is largely a molecule-specific parameter that determines the shape of the ionization probability as a function of orientation angle.

We note that in Eq. \eqref{eq:asymptotic_coeff0}, the orientation dependence is encoded in the orientation of $\big| \Psi_n^{(0)} \big\rangle$, $\big| \Psi_{n'}^{+(0)} \big\rangle$, and $\hat{V}_C$ relative to $\Omega_\nu^{n'n}(\mathbf r) = R_\nu^{n'n(0)}(\eta) \, \Phi_\nu^{n'n(0)}(\xi,\varphi) / \sqrt{\eta}$, which is fixed in space since it is contingent to the lab-frame $z-$axis. This means, for every orientation angle of interest, one needs to rotate the $3N$ and $3(N-1)$ dimensional wave functions and potential. A more efficient but equivalent method is to fix the wave functions and potential, but rotate $\Omega_\nu^{n'n}(\mathbf r)$ instead. To perform this three-dimensional rotation, we follow a partial wave expansion method employed in \cite{ir_oewfat_grid}
\begin{align}
    \label{eq:asymptotic_coeff1}
    g_{\nu\sigma}^{n'n}(\beta,\gamma) = 
    \sum_{l=|m|}^\infty  \sum_{m'=-l}^l
    I_{\nu lm'\sigma}^{n'n} \, d_{mm'}^l(\beta) \, e^{-im'\gamma}
\end{align}
where
\begin{align}
    \label{eq:mewfat_integral1}
    I_{\nu lm'\sigma}^{n'n} 
    =&\,\, 
    \int dQ_N \,\,
    \tilde{R}_{\nu l}^{n'n}(r)
    Y_{lm'}^*(\theta,\varphi) \,
    \chi_{\sigma}^\dagger(s) \nonumber\\
    &\,\,
    \times
    \left(\Psi_{n'}^{+(0)}(Q_{N-1})\right)^*
    V_c(\mathbf R_N) \Psi_n^{(0)}(Q_N),   \\
    \tilde R_{\nu l}^{n'n}(r) 
    =& \,\, 
    \omega_{\nu l}^{n'n}
    (\varkappa r)^l
    e^{-\varkappa r}
    M(l+1-Z/\varkappa, 2l+2, 2\varkappa r),  \nonumber
\end{align}
$d_{mm'}^l(\beta)$ is the Wigner function \cite{quantum_angular_momentum}, $Y_{lm}(\theta,\varphi)$ is the spherical harmonics, $M(a,b,x)$ is the confluent hypergeometric function \cite{math_handbook_abramowitz}, and
\begin{align*}
    \omega_{\nu l}^{n'n} 
    =& \,\,
    (-1)^{l+(|m|-m)/2+1} \,
    2^{l+3/2} \,
    \varkappa^{Z/\varkappa - (|m|+1)/2 - n_\xi}   \nonumber \\
    &\,\,
    \times
    \sqrt{(2l+1) (l+m)! (l-m)! (|m|+n_\xi)! n_\xi!}   \nonumber \\
    &\,\,
    \times
    \frac{l!}{(2l+1)!} 
    \sum_{k=0}^{\textrm{min}(n_\xi,l-|m|)}
    \frac
    {1}
    {k! (l-k)! (|m|+k)!}   \nonumber \\
    &\,\, 
    \times
    \frac{\Gamma(l+1-Z/\varkappa+n_\xi-k)}{(l-|m|-k) !(n_\xi-k)!}.
\end{align*}
is a normalization coefficient. 
To compute the integral in Eq. \eqref{eq:mewfat_integral1}, the integration over $Q_{N-1}$ and $s_N$ is performed analytically, while the integration over $\mathbf r_N$ is done with the help of a combination of Becke fuzzy cells \cite{becke_fuzzy_cell} and quadrature grids \cite{ir_oewfat_grid}.

As has been mentioned earlier, the WFAT ionization probability is origin-independent provided that the asymptotic behavior of the wave functions is exact. For all simulations presented in this work, we use Gaussian-type orbitals (GTO), which are excellent for representing local properties but are poor for simulating asymptotic behavior. For this reason, the origin-independence in GTO-based WFAT results is only approximate. We can, however, minimize this effect by following the previous works (\textit{e.g.} see Ref. \cite{ir_oewfat_hf}) where the origin is chosen so that 
\begin{align}
    \mathbf D_{n'}^{(N-1)} =& \, 
    \sum_{I=1}^{N_A} Z_I \mathbf C_I -
    \boldsymbol \mu_{n'}^{(N-1)}
    = 0,
\end{align}
here, $\mathbf D_{n'}^{(N-1)}$ is the total dipole of the final charge state.
All simulations presented here are performed in the frame where $\mathbf D_{n'}^{(N-1)}=0$.

Concluding this section, we briefly discuss the mathematical structure of the expressions for the asymptotic coefficient in IR ME-WFAT and TR ME-WFAT. First, in the case of OE-WFAT, both the IR (see Eq. (20) in Ref. \cite{ir_oewfat}) and the TR (see Eq. (59) in Ref. \cite{ir_oewfat}) formulas explicitly show a dependence on the ionizing orbital, which is one of the initial charge state orbitals.
In ME-WFAT, the ionizing orbital is the Dyson orbital. The dependence of the asymptotic coefficient on the orbital is explicit in TR ME-WFAT (see Eq. \eqref{eq:asymptotic_coeff_tr_mewfat} below)
while in IR ME-WFAT, it is partially explicit (see Eq. \eqref{eq:mewfat_integral1}). It is only partially explicit because as will
be shown in Section \ref{sec:HF_application}, the 1-electron term, $V_{1e}(\mathbf r)$, will lead to the appearance of the Dyson orbital, but the 2-electron term, $V_{2e}(\mathbf R_N)$, will not. Despite this property, we will show in Section \ref{sec:HF_application} that the present formulation of ME-WFAT in the IR is consistent with that in the TR.

\subsection{The asymptotic cancellation problem \label{sec:asymptote_cancel}}
The hallmark of TR WFAT (whether one- or many-electron) is the exact cancellation in the asymptotic region between a certain exponential pre-factor and another function that depends on the ionizing orbital. For this reason, the asymptotic part of the ionizing orbital needs to be as close as possible to being the inverse of the 
exponential pre-factor, 
For instance, in the TR ME-WFAT, the structure factor is given by
\begin{align}
    \label{eq:asymptotic_coeff_tr_mewfat}
    g_{\nu\sigma}^{n'n} 
    =& \,
   \Bigg[
    \eta^{1/2-\beta_\nu^{n'n(0)}/\varkappa}
    e^{\varkappa \eta/2}
    \int_0^\infty \int_0^{2\pi} 
    d\xi d\varphi \,
    \phi_\nu^{n'n(0)}(\xi) \nonumber \\
    & \times
    \frac{e^{im\varphi}}{\sqrt{2\pi}}
    \psi_D^\sigma(\mathbf r)
    \Bigg]
    \Bigg|_{\eta \to \infty}
\end{align}
\cite{tr_mewfat} where 
\begin{align*}
    \psi_D^\sigma (\mathbf r) 
    =& \, 
    \int dQ_{N-1} ds \,
    \left( \Psi_{n'}^{+(0)}(Q_{N-1}) \, \chi_\sigma(s) \right)^* 
    \Psi_n^{(0)}(Q_N)
\end{align*}
is the Dyson orbital. In the above equation, the integral over $s$ is to be understood as a vector scalar product. The exact asymptotic behavior of the Dyson orbital is that it decays as $\exp{-\varkappa \eta/2}$ \cite{tr_mewfat_app2}, hence asymptotically cancelling the exponential function in Eq. \eqref{eq:asymptotic_coeff_tr_mewfat}. For approximate many-electron initial and/or final wave functions, however, the exponent $\tilde\varkappa$ appearing in the asymptotic tail of the corresponding Dyson orbital is not necessarily equal to $\varkappa$. This may cause $g_{\nu\sigma}^{n'n}$ either to vanish ($\tilde\varkappa > \varkappa$) or to diverge ($\tilde\varkappa < \varkappa$). 

The above cancellation problem has been discussed in Ref. \cite{tr_mewfat_app2}. There, 
the many-electron wave functions are calculated at the Hartree-Fock level, in which the Dyson orbital is just a linear combination of the occupied orbitals. The asymptotic part of such a linear combination is therefore given by that of the least bound orbital, which has the largest
coefficient, \textit{i.e.} $\tilde\varkappa = \sqrt{2|\epsilon|} \neq \varkappa$ where $\epsilon$ is the orbital energy of the least bound occupied orbital. To ensure that a finite value of the asymptotic coefficient can be extracted, Tolstikhina \textit{et al.} \cite{tr_mewfat_app2} imposes $\tilde{\varkappa} = \sqrt{2 |\epsilon|} = \varkappa$.
The situation becomes more complicated when a configuration interaction (CI) wave function is used since in this case, orbitals having lower binding energy appear in the Dyson orbital linear combination. These orbitals decay slower than HF HOMO, which causes an imperfect cancellation in the asymptotic region.

In IR ME-WFAT, the equation used to calculate the asymptotic coefficient (Eq. \eqref{eq:asymptotic_coeff0} or \eqref{eq:asymptotic_coeff1}) involves integration over the entire $4N$ space-spin coordinates, in contrast to the TR ME-WFAT version in Eq. \eqref{eq:asymptotic_coeff_tr_mewfat} that leaves $\eta$ unintegrated. The fact that these integrals in IR ME-WFAT are over all space means that the asymptotic part of the Dyson orbital is no longer crucial in determining the convergence of the integration, which further implies that Gaussian standard basis functions can be used to span the orbital space making up the multielectron wave functions. It is also for the same reason that, when the one-electron approximation is sufficient, the IR OE-WFAT \cite{ir_oewfat, ir_oewfat_hf, ir_oewfat_grid} is more practical than the TR version for polyatomic molecular tunnel ionization problems. Lastly, although it is still a topic for future studies, the use of CI wave functions will be much more manageable in IR ME-WFAT compared to TR ME-WFAT \cite{tr_mewfat_app2}.


\subsection{Ionization rate}
The difference between TR and IR lies in the method to calculate the asymptotic coefficient, while the fundamental assumption on how the ionization probability is extracted remains unchanged. Thus, the ionization rate formula is the same as in TR \cite{tr_mewfat}, namely
\begin{align}
    \label{eq:ion_rate}
    \Gamma = \sum_{n'\sigma\nu} N|f_{\nu\sigma}^{n'n(0)}|^2 + O(\Gamma^2).
\end{align}
Eq. \eqref{eq:ion_rate} above gives the ionization rate of an electron originating from the initial state $|\Psi_n\rangle$ to a number of final states $\left| \Psi_{n'}^+ \right\rangle$ having one less electron. The presence of the multiplier $N$ in Eq. \eqref{eq:ion_rate} accounts for the indistinguishability of the ionized electrons.

\subsection{Application to single-determinant wave functions} \label{sec:HF_application}

It is important to note that no assumptions are imposed on the unperturbed wave functions $|\Psi_n\rangle$ and $\big| \Psi_{n'}^+ \big\rangle$ appearing in Eq. \eqref{eq:mewfat_integral1} other than being an eigenstate of the respective all-electron Hamiltonian---both can take any structure, whether single (Hartree-Fock or DFT \cite{wahyutama_irmewfat_dft}) or multideterminant. Since determinants are the building blocks of many correlated wave function 
methods, in this section, IR ME-WFAT will be applied to the calculation of the asymptotic coefficient when both the initial and final wave functions are a single (Slater) determinant. 

In this case, we have
\begin{subequations}
   \label{eq:slater_det}
   \begin{align}
       \Psi(Q_N) =& \,
       \frac{1}{\sqrt{N!}}
       \operatorname{det}\left(
       \psi_{i_1}^\alpha \cdots \psi_{i_{N_\alpha}}^\alpha \, 
       \psi_{i_1}^\beta \cdots \psi_{i_{N_\beta}}^\beta
       \right), \\
       \Psi^+(Q_{N-1}) =& \,
       \frac{1}{\sqrt{(N-1)!}}
       \operatorname{det}\left(
       \upsilon_{i_1}^\alpha \cdots \upsilon_{i_{N'_\alpha}}^\alpha \, 
       \upsilon_{i_1}^\beta \cdots \upsilon_{i_{N'_\beta}}^\beta \right)
   \end{align}
\end{subequations}
where $N=N_\alpha + N_\beta$, $N-1=N'_\alpha + N'_\beta$, and we have omitted the eigenstate index from the two wave functions since the HF method produces only one meaningful determinant. 
Each spin orbital is a product state between the spatial and spin functions, \textit{e.g.} $\upsilon_i^\beta \equiv \upsilon_i^\beta(\mathbf r) \chi_\beta(s)$. Since the ionized electron is assumed to have a definite $z$-component of spin, we have
$N'_{\sigma} = N_{\sigma} - 1$ and $N'_{p(\sigma)} = N_{p(\sigma)}$.
Here, the spin channel where an electron has been removed (that is, the \textit{ionized spin channel}) is denoted by $\sigma$,
while $p(\sigma)$ is the \textit{unionized spin channel} and has values such that $p(\alpha)=\beta$ and $p(\beta)=\alpha$.

We note that with the help of Eq. \eqref{eq:vc_v1e_v2e}, 
$I_{\nu lm'\sigma}^{n'n}$ can be decomposed into two- and one-electron terms,
\begin{gather}
    I_{\nu lm'\sigma}^{n'n} = 
    \left\langle \Psi^+; \Omega_{lm'}^\nu \chi_{\sigma} 
    \left| 
    \hat V_{1e}
    \right|
    \Psi \right\rangle
    +
    \left\langle \Psi^+;\Omega_{lm'}^\nu \chi_{\sigma} \left| \hat V_{2e}    \right| \Psi \right\rangle 
\end{gather}
Then, the integrals over the first $N-1$ electron coordinates as well as $s_N$ are carried out analytically while the one over $\mathbf r_N$ is done numerically using numerical quadrature.
The former involves extensive determinant algebra, and thus we will only cite the results of Appendix \ref{sec:HF_integral_derive} here. By first defining $\Omega_{lm'}^\nu(\mathbf r) = \tilde{R}_{\nu l}(r) Y_{lm'}(\theta,\varphi)$, the two-electron component reads
   \begin{align}
       \label{eq:v2e_term}
       \left\langle \Psi^+;\Omega_{lm'}^\nu   \right. &  \left. \chi_{\sigma} \left| \hat V_{2e}    \right| \Psi \right\rangle 
       = \nonumber \\
       &\,\frac{\delta_{M_s' + m_{\sigma}, M_s}}{\sqrt{N}} \,    (-1)^{N+\delta_{\sigma\beta}N_\alpha}  \nonumber \\
       &
       \times \Bigg\{  
       \mathcal{R} \sum_{k'=1}^{N_{\sigma}-1} \, \sum_{j=2}^{N_{\sigma}} \,    \sum_{k=1}^{j-1} (-1)^{j+k+k'} \mathcal{Q}(k,j,k') 
       \nonumber \\
       &
       \times
       \left( \left\langle \Omega_{lm'}^\nu \left| \hat V_{k'k}^{\sigma} \right|    \psi_j^{\sigma} \right\rangle   
       -\left\langle \Omega_{lm'}^\nu \left| \hat V_{k'j}^{\sigma} \right|    \psi_k^{\sigma} \right\rangle \right) 
       + \nonumber \\
       &
       \big\langle \Omega_{lm'}^\nu \big| \hat{\mathcal{V}}^{\sigma} \big|    \tilde{\psi}^{\sigma} \big\rangle 
       \Bigg\}
   \end{align}
and for the 1-electron term, it reads
\begin{align}
    \label{eq:v1e_term}
    \left\langle \Psi^+; \Omega_{lm'}^\nu \chi_{\sigma} 
    \left| 
    \hat V_{1e}
    \right|
    \Psi \right\rangle = 
    \delta_{M_s'+m_{\sigma},M_s} \left\langle \Omega_{lm'}^\nu \, \Big| \hat V_{1e} \Big| \psi_D^{\sigma} \right\rangle.
\end{align}
where $\mathcal{Q}(k,j,k')$ is the determinant of the overlap matrix in the \textit{ionized spin channel} between the initial wave function after removing $\{\psi_k^\sigma, \psi_j^\sigma\}$ and the final wave function after removing $\{\upsilon_{k'}^\sigma\}$, $\mathcal{R}$ is the determinant of the overlap matrix in the \textit{un-ionized} 
spin channel between the initial and final eigenstates,
$\hat V_{k'k}^\sigma$ is the electrostatic repulsion operator due to the $(\upsilon_{k'}^\sigma(\mathbf r))^* \, \psi_k^\sigma(\mathbf r)$ charge density, and $\hat{\mathcal V}^\sigma$ is the repulsion operator of the electrons in the spin-$p(\sigma)$ channel. For more details, see Appendix \ref{sec:HF_integral_derive}. In Eq. \eqref{eq:v2e_term}, $k'$ runs over the occupied spatial orbitals in the spin-$\sigma$ channel of the final state whereas $j$ and $k$ run over the occupied spatial orbitals in the spin-$\sigma$ channel of the initial state.

A brief discussion of the various terms appearing in the RHS of Eq. \eqref{eq:v2e_term} is in order. First, the first and second terms inside the curly bracket represent the electron-electron interaction (repulsion and exchange) 
felt by the ionized electron in its own spin channel (spin-$\sigma$ channel). The third term may be viewed as the repulsion 
due to the electrons in the spin-$p(\sigma)$ channel felt by the ionized electron. There is no cross-channel exchange term, that is, the exchange between the ionized electron and those in the spin-$p(\sigma)$ channel, which is a well-known property associated with the use of single-determinant wave functions.


We are now at the point where we can show that IR ME-WFAT 
is equivalent to IR OE-WFAT in a certain limit. This is when all of the occupied spin orbitals of the final HF wave function are a subset of the occupied spin orbitals of the initial HF wave function, in other words when the final cationic determinant is unrelaxed. If we assume that the only spin orbital that is unoccupied in the final HF wave function is $|\psi_{i'}^{\sigma} \rangle$, then the determinant coefficients defined in Eq. \eqref{eq:det_coeff_general} reduce to
\begin{subequations}
   \label{eq:det_coeff_unrelaxed}
   \begin{align}
       \mathcal P(i) &= \delta_{ii'} \\
       \mathcal R &= 1 \\
       \mathcal Q(k,i,k') &= (1-\delta_{ik})(\delta_{ii'} \delta_{k'\Bar k} + \delta_{ki'} \delta_{k' \Bar i}) \\
       \mathcal S(k,k') &= \delta_{k'k}
   \end{align}
\end{subequations}
where $\Bar i = i$ if $i<i'$ and $\Bar i = i-1$ otherwise.
$\Bar k$ has the same definition as $\Bar i$ with all occurrences of $i$ replaced by $k$, except for $i'$. 
Then, by Koopman's theorem, the ionization potential is equal to the orbital energy of $|\psi_{i'}^{\sigma}\rangle$. The electronic dipole moment difference is then also equal to the dipole moment of $|\psi_{i'}^{\sigma}\rangle$. As is shown in Appendix \ref{sec:oewfat_from_mewfat}, by putting Eq. \eqref{eq:det_coeff_unrelaxed} into Eq. \eqref{eq:v1e_term} and \eqref{eq:v2e_term}, one recovers the integral terms valid for IR OE-WFAT (\textit{e.g.} see Eq. (31), (21), (32)-(34) in Ref. \cite{ir_oewfat_hf}).
Concluding this section, we have obtained the integral formulas necessary to calculate ME-WFAT ionization rate in the integral representation when the initial and final wave functions are both single Slater determinants. 

\section{Results and Discussion}  \label{sec:results}
In TR ME-WFAT, it is clear that the Dyson orbital is central in determining the angular dependence of the structure factor (see Eq. \eqref{eq:asymptotic_coeff_tr_mewfat}). On the other hand, in IR ME-WFAT, the significance of this orbital may not be directly apparent (cf. Eq. \eqref{eq:asymptotic_coeff0}). Reflecting 
the central role played by the Dyson orbital in ME-WFAT, we use it as one of the components of analyses presented in this section. In this regard, we introduce the ratio 
\begin{align}
    \label{eq:r_ratio}
    r = \left| \frac{\mathcal P(i_{\textrm{2nd max}})}{\mathcal P(i_{\textrm{max}})} \right|,
\end{align}
where $i_{\textrm{max}}$ and $i_{\textrm{2nd max}}$ are the indices of the largest and second-largest expansion coefficients of the Dyson orbital, respectively (see Eq. \eqref{eq:dyson_define} and \eqref{eq:incomplete_dyson}). This ratio may be interpreted as the degree of similarity of the Dyson orbital to one of the occupied MO of the initial state.
\rladd{For general multiconfiguration wave functions, two factors influence $r$: (1) relaxation effect, and (2) multireference effect. In the present work where all states are single-determinant, only (1) applies.}
The other component of our analysis is the dipole moment because the dipole moments used in OE-WFAT and in ME-WFAT may have different directions and as will be shown below, this affects the relative magnitude of the ionization rates between the parallel and antiparallel directions. 

\rladd{Some of the systems considered below are doubly degenerate, these are O$_2$, OCS, and CH$_3$F. In the exact treatment, their degeneracies will be broken by the ionizing field, however, in the leading order approximation of WFAT assumed in this work, only field-free wave functions and orbitals are used with any degeneracies still intact. It is only when one goes beyond the leading order approximation, should the field-induced modification on the orbitals be taken into account, possibly removing any degeneracies \cite{wfat3, tr_mewfat_first_order}.}
\rlremove{In what follows, before presenting the results for molecules, we will first discuss the equivalence between TR and IR ME-WFAT in Section \ref{sec:atoms} using several atoms.} 
The neutral and cation wave functions employed in all of the results \rlremove{presented here} are obtained from a HF calculation, \rladd{thus representing a ground-state-to-ground-state ionization channel}, and the formulas derived in Section \ref{sec:HF_application} have been used.

\subsection{Atoms} \label{sec:atoms}
In this section, we will first validate IR ME-WFAT  by comparing atomic asymptotic coefficients with those obtained by TR ME-WFAT in Ref. \cite{tr_mewfat_app2}. The asymptotic coefficients of He, Li, Be, Na, and Mg calculated at the HF level for
ionization starting from the neutral and ending up in the cation are presented in Table \ref{tab:atoms}. In the table, $p$ is a constant pre-factor arising from the fact that in HF, one can ionize an electron with two (one) possible spin orientations when the neutral has no (one) unpaired electrons. $g_{00}$ in Table \ref{tab:atoms} is obtained by multiplying the asymptotic coefficient calculated by Eq. \eqref{eq:asymptotic_coeff0} (or Eq. \eqref{eq:asymptotic_coeff1}) with $\sqrt{p}$, while $R=\mathcal{P}(i_\textrm{max}) \mathcal{R}$ and $\varkappa$ is given by Eq. \eqref{eq:wronskian}. $R$ may be considered as a measure of the influence of relaxation on the cation wave function, with the maximum value of unity representing the situation where the final state is unrelaxed. $g_{00}$ constitutes the result of IR ME-WFAT as formulated in the present work. We compare $g_{00}$ with $g_\textrm{ref}$, which is the TR ME-WFAT asymptotic coefficient extracted from Ref. \cite{tr_mewfat_app2}. The agreement is generally promising with the largest difference being about $6\%$ found for Mg among the atoms simulated.

\begin{table}
   \caption{\label{tab:atoms} The asymptotic coefficients of He,  Li, Be, Na, and Mg calculated by IR ME-WFAT ($g_{00}$ and $\tilde g_{00}$) and TR ME-WFAT ($g_\textrm{ref}$). aug-pc-3 basis were employed for these simulations except for Li where cc-pvtz is used. This is because HF calculation of Li using aug-pc-3 gives wrong valence orbital ordering.}
   \begin{ruledtabular}
      \begin{tabular}{lrrrrrrr}
         & & & & & & & \\ [-0.8em]
         Atom &
         $p$ &
         $R$ &
         $\varkappa$ &
         $g_{00}$\footnotemark[1] &
         $\tilde{\varkappa}$ &
         $\tilde{g}_{00}$\footnotemark[2] &
         $g_\textrm{ref}$\footnotemark[3]   \\
         & & & & & & & \\ [-0.8em]
         \colrule
         &&&&&&& \\ [-0.6em]
         He   & 2  &  0.984  &  1.312 78  & 2.885  & 1.354 96  & 2.626  & 2.942 \\
         &&&&&&& \\ [-0.9em]
         Li   & 1  &  1.000  &  0.626 61  & 0.474  & 0.626 66  & 0.474  & 0.474 \\
         &&&&&&& \\ [-0.9em]
         Be   & 2  &  0.984  &  0.768 84  & 1.324  & 0.786 46  & 1.437  & 1.392 \\
         &&&&&&& \\ [-0.9em]
         Na   & 1  &  1.000  &  0.603 23  & 0.416  & 0.603 58  & 0.416  & 0.415 \\
         &&&&&&& \\ [-0.9em]
         Mg   & 2  &  0.985  &  0.696 75  & 1.022  & 0.711 32  & 1.068  & 1.087 \\
      \end{tabular}
   \end{ruledtabular}
   \footnotetext[1]{This work, using the proper value of $\varkappa$ given by Eq. \eqref{eq:wronskian}.}
   \footnotetext[2]{This work, replacing the field-free ionization potential in Eq. \eqref{eq:wronskian} with $\epsilon$, the orbital energy of the least bound occupied orbital.}
   \footnotetext[3]{The value calculated in Ref. \cite{tr_mewfat_app2} using TR ME-WFAT.}
\end{table}

The values of $g_\textrm{ref}$ in Table \ref{tab:atoms} were calculated by using $\epsilon$, the orbital energy of the least bound occupied orbital instead of the actual ionization potential in Eq. \eqref{eq:wronskian} ($\varkappa \Rightarrow \tilde\varkappa = \sqrt{2|\epsilon|}$) \cite{tr_mewfat_app2}. One may therefore speculate that if we do the same substitution wherever $\varkappa$ appears in the integral formulation derived in this work, then the agreement may get better. The results of this substitution of ionization potential with the least bound orbital energy are shown in the \nth{6} and \nth{7} columns of Table \ref{tab:atoms}. As one can see, for He, Be, and Mg, the agreement between $\tilde g_{00}$ and the reference value improves as one goes down the table with the relative errors of $10.7\%$, $3.3\%$, and $1.8\%$, respectively. On the other hand the agreement between $g_{00}$ with the reference value gets worse along the same direction with the relative errors of $1.9\%$, $4.9\%$, and $6.0\%$. This shows that the substitution $\varkappa \Rightarrow \tilde\varkappa$ does not guarantee improvement of the agreement between IR ME-WFAT and TR ME-WFAT \cite{tr_mewfat_app2}, which further implies that the results presented in this section (as well as those that follow) are not totally identical to what they would be if they had been obtained through the tail representation formulation.
We identify the source of this imperfect equivalence as being due to the level of approximation to the all-electron wave functions involved in the calculation (Hartree-Fock level). In TR ME-WFAT, the information on the cation wave function enters solely through the Dyson orbital (see Eq. \eqref{eq:asymptotic_coeff_tr_mewfat}), whereas in IR ME-WFAT, apart from the Dyson orbital, it also enters through the matrix element of the one-body Coulomb potential (the $\hat V_{k'j}^\sigma$ terms in Eq. \eqref{eq:v2e_term}). To reconcile this difference in the mathematical structure of the two representations, it is conceivable that the level of approximations to the all-electron wave functions should play a role in bringing the equivalence closer or farther away. We would like to emphasize, however, that this inequivalence will disappear as one uses more and more accurate all-electron wave functions for the neutral and cation.


In Table \ref{tab:atoms}, Li and Na present an exception in which their $g_{00}$ and $g_\textrm{ref}$ are essentially identical, although HF wave functions are used. The reason is that, as can be seen from their values of $R$, which is unity, the HF procedure for the cation of these atoms produces orbitals that are almost identical to those making up the neutral HF wave function, that is, the cation is almost unrelaxed. In that case, both TR and IR ME-WFAT reduces to OE-WFAT.

\subsection{O$_2$} \label{sec:o2_no}
\begin{figure}
    \includegraphics[width=1\columnwidth]{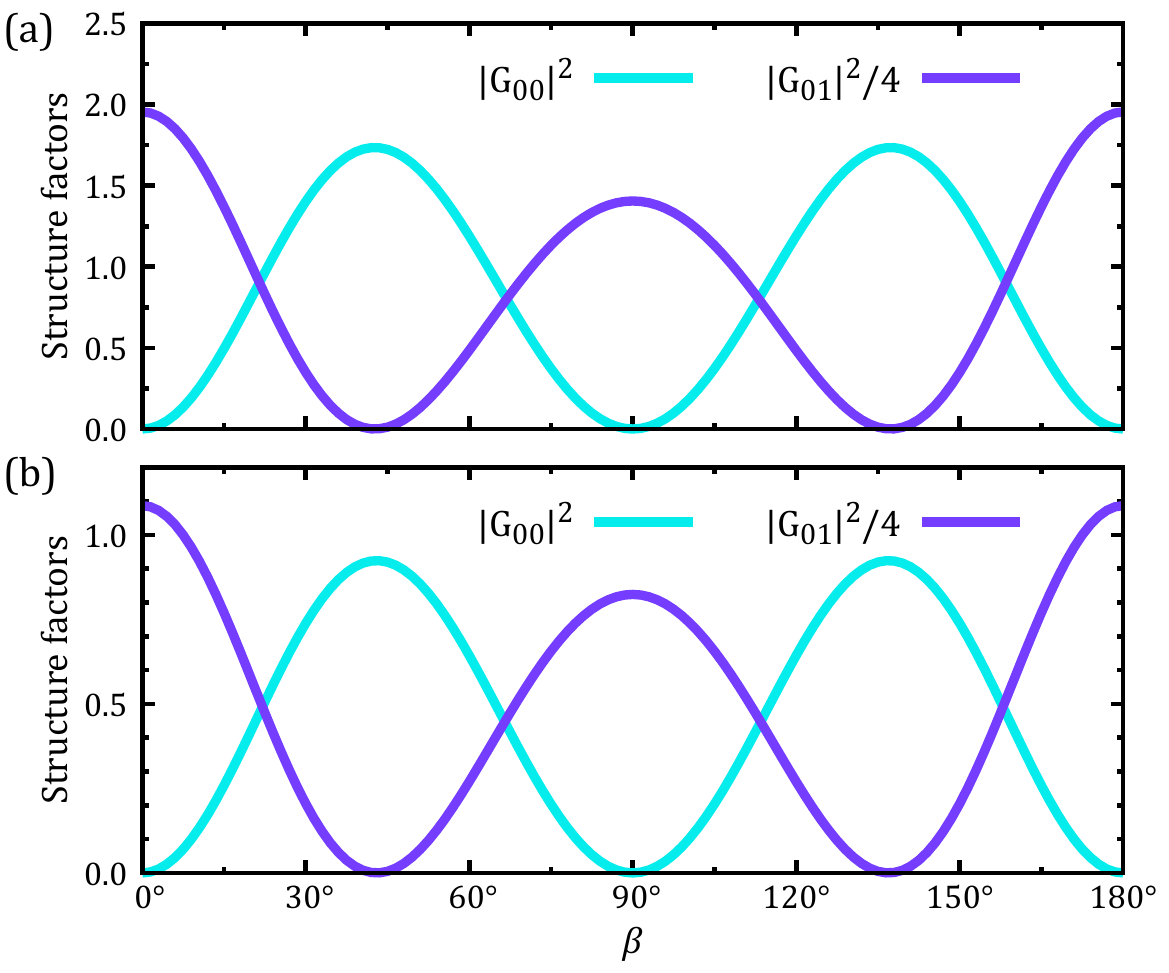}
    \caption{\label{fig:o2} Structure factors of O$_2$ simulated using (a) OE-WFAT and (b) ME-WFAT. In (a), HOMO from Hartree-Fock calculation of O$_2$ is used. The structure factors in (a) and (b) are a slice at a constant $\gamma$, which is the direction of the lobe of HOMO and Dyson orbital density, respectively. When $\beta=0^\circ$, the field is parallel with the molecular axis.}
\end{figure}
For single configuration wave functions, OE-WFAT is a special case of ME-WFAT where the cation configuration is formed out of a subset of $N-1$ orbitals of the occupied orbitals of the neutral. It should then be expected that in molecules where this condition is nearly satisfied, the structure factors calculated by OE-WFAT and ME-WFAT should be similar. In this section, we show that O$_2$ falls within this group of molecules.

Fig. \ref{fig:o2}(a) shows the structure factor for the parabolic channel $(n,m)=(0,0)$ and $(n,m)=(0,1)$ of O$_2$ calculated using OE-WFAT and a bond length of $1.158$ \AA. The orbital used for this calculation is the HOMO, which is doubly degenerate and has a $\pi_g$ symmetry. The symmetry is reflected in the shape of the structure factor $|G_{00}|^2$ exhibiting two peaks at $\beta=42^\circ$ and $\beta=138^\circ$, and \rladd{a mirror} symmetry
around $\beta=90^\circ$. The ionization yield of O$_2$ interacting with an oscillating electric field of low intensity simulated using TDCI in Ref. \cite{schlegel_o2_tdci-2015} has a very similar double-peak structure located at $45^\circ$ and $135^\circ$. The shape as well as the position of these peaks also agrees with the experiments on O$_2$ reported in Ref. \cite{o2_experiment}. 
The ME-WFAT results shown in Fig. \ref{fig:o2}(b) are very similar to the one-electron counterpart in panel (a) for both parabolic channels shown, with the only difference being the magnitude. In fact, the $r$ ratio for O$_2$ obtained by HF is zero, with the largest coefficient belonging to HOMO, making the O$_2$ Dyson orbital essentially the HOMO itself. The basis set, the ratio $r$, the largest and second-largest contributors to the Dyson orbital, and the ionization potentials used in ME-WFAT or OE-WFAT for O$_2$ as well as other molecules studied here are summarized in Table \ref{tab:properties}. 

\rladd{Angle-dependent ionization yields of O$_2$ have also been studied in the context of field-induced multielectron polarization effects (MEP) in Ref. \cite{multielect-nonpolar-2022}. It was found that MEP has negligible effects on the angle-dependent yield of O$_2$. Although MEP is not considered in this work---remember that only field-free orbitals enter the formulation, this observation agrees with ours in that multielectron effects are negligible in the ionization from the ground state of O$_2$ to that of its cation. To formally incorporate MEP in WFAT framework, one must include the 1st order correction of ME-WFAT, which, however, lies outside the scope of the present work, but will be studied in a future publication.
}

\begin{table*}
   \caption{\label{tab:properties} The basis set, the $r$ ratio (Eq. \eqref{eq:r_ratio}), the largest and 2nd largest neutral's MO, and ionization potential (IP) of the molecules studied in this work. The IP's are in hartree, and the values to the left and right of the slash sign is the IP for ME-WFAT and OE-WFAT, respectively.}
   \begin{ruledtabular}
      \begin{tabular}{ccrrrr}
         & & & & & \\[-0.9em]
         Species &
         Basis set &
         $r$ &
         $i_{\textrm{max}}$ &
         $i_{\textrm{2nd max}}$ &
         $\textrm{IP}^{(0)}$ \\
         & & & & & \\[-0.9em]
         \hline
         & & & & & \\[-0.8em]
          O$_2$   & aug-cc-pvqz & $0.0$    & HOMO   & --     & $-0.470329$ / $-0.537404$ \\
          OCS     & aug-cc-pvqz & $0.050$  & HOMO   & HOMO-2 & $-0.369409$ / $-0.416889$ \\
          CO      & aug-cc-pvqz & $0.114$  & HOMO   & HOMO-2 & $-0.478264$ / $-0.551712$ \\
          N$_2$   & aug-cc-pvqz & $0.030$  & HOMO-1 & HOMO-3 & $-0.578156$ / $-0.628166$ \\
          HCOOH   & pc-3        & $0.198$  & HOMO   & HOMO-5 & $-0.368459$ / $-0.474097$\\
          CH$_3$F & pc-3        & $0.588$  & HOMO   & HOMO-2 & $-0.455863$ / $-0.529495$\\
      \end{tabular}
   \end{ruledtabular}
\end{table*}

\subsection{OCS and CO} \label{sec:co_ocs}

As mentioned in Section \ref{sec:intro}, there are two main reasons why 
the shape of the angle-dependent ionization probability obtained by ME-WFAT 
may differ substantially from OE-WFAT: (i) the dipole moment and (ii) the dissimilarity between the HOMO and the Dyson orbital. The present and the next sections are devoted to illustrating the effect of each of these factors on the angular variation of the structure factor.

\begin{figure}
    \includegraphics[width=\columnwidth]{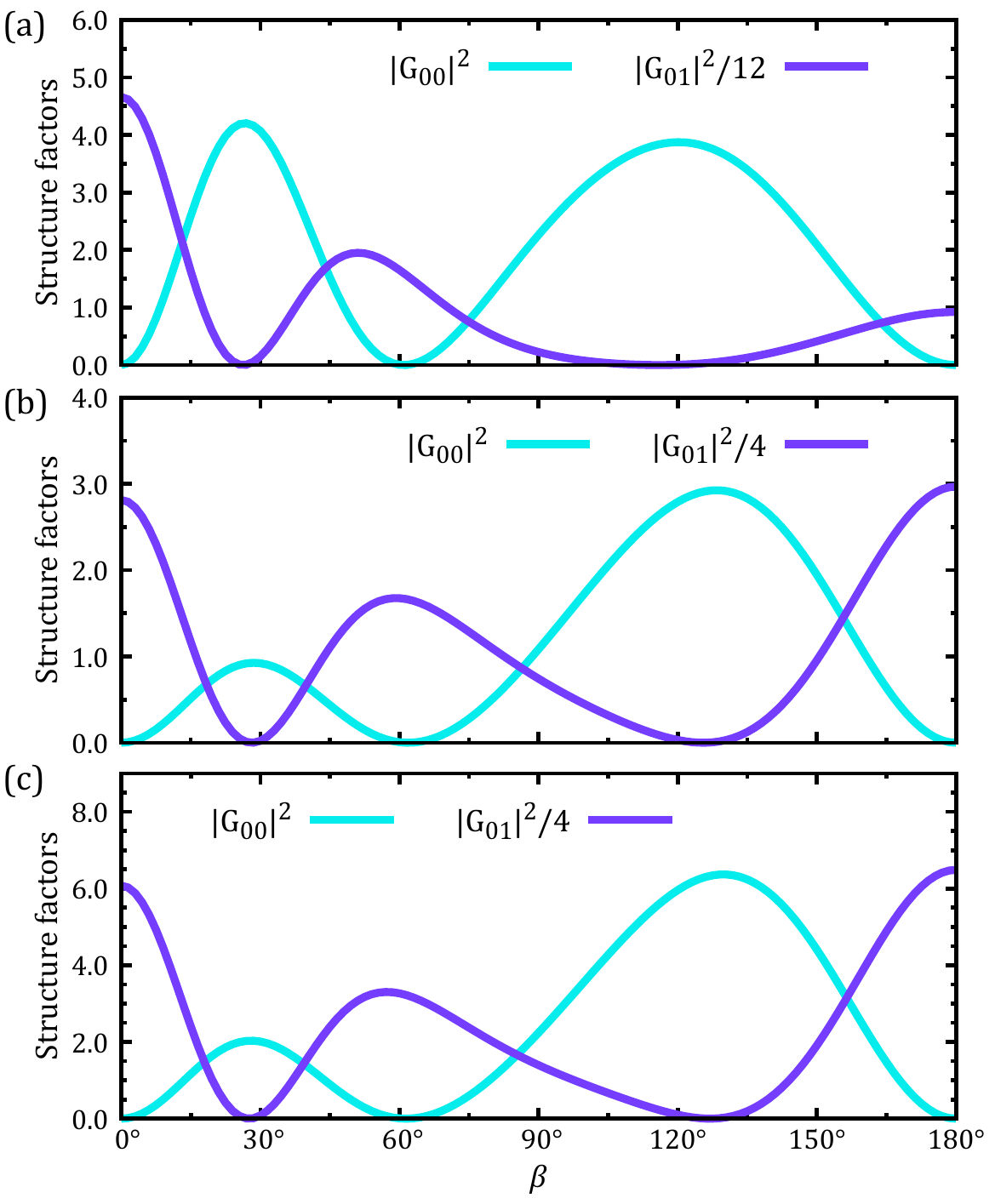}
    \caption{\label{fig:ocs} Structure factors of OCS simulated using (a) OE-WFAT and (b) ME-WFAT. In (a), HOMO from Hartree-Fock calculation of OCS is used. (c) The structure factors calculated with OE-WFAT but where the definitions for the dipole moment and origin use those from ME-WFAT. The structure factors in (a), (b), and (c) 
    are a slice at a constant $\gamma$, which is the direction of the lobe of the ionizing orbital (the Dyson orbital in ME-WFAT or the HOMO in OE-WFAT) density.}
\end{figure}
We first look at OCS, whose HOMO has a $\pi$ symmetry and is doubly degenerate. 
\rladd{OCS is one of the most frequently studied molecules for its strong-field ionization properties \cite{ocs-align-midir-2016, ocs-single-double-2018, ocs-sfi-2020, wahyutama_irmewfat_dft, ocs-ac-stark-2023}.}
In our simulations, we choose the O--C and C--S bond lengths to be $1.123$ \AA~and $1.564$ \AA, respectively.
The double-peak structure of $|G_{00}|^2$ obtained by OE-WFAT seen in Fig. \ref{fig:ocs}(a) is a manifestation of the $\pi$ symmetry of the HOMO.
The structure factors obtained by ME-WFAT (Fig. \ref{fig:ocs}(b)) still retain the same general features as OE-WFAT (Fig. \ref{fig:ocs}(a)) for both parabolic channels simulated here but this time the magnitude of the peak of the $G_{00}$ channel around $\beta\sim 125^\circ$ is three times larger than the other peak. 
Here, we will show that this difference is largely due to the different treatment of the dipole moment in the two methods. We run a OE-WFAT simulation but this time, the origin and dipole moment are chosen according to ME-WFAT prescription. This means that the origin is such that $\mathbf D^{(N-1)} = 0$ and the dipole moment is given in the explanation immediately following Eq. \eqref{eq:cation_emoment}.
The rest, such as the orbital and integral formulas, are the same as those used to obtain Fig. \ref{fig:ocs}(a). The structure factors resulting from these modifications to OE-WFAT are shown in Fig. \ref{fig:ocs}(c). We see that in the case of OCS, mixing the OE-WFAT prescription for orbital and integral formula with the ME-WFAT prescription for the origin and dipole moment closely recovers the shape of the structure factors obtained by ME-WFAT. 

\rlremove{We note that the angle-dependent ionization yield for OCS obtained using DFT-based IR ME-WFAT has been calculated in Ref. \cite{wahyutama_irmewfat_dft}, and it agrees with $|G_{00}|^2$ in Fig. \ref{fig:ocs}(b).}
\rladd{We note that alignment-dependent rates, rather than orientation-dependent ones, of OCS seem to be more commonly calculated or measured \cite{ yield-asym-top-2011, ocs-align-midir-2016, ocs-single-double-2018, ocs-sfi-2020}. In these alignment-dependent curves, the maximum occurs at $\beta=90^\circ$. Approximately emulating an alignment-dependent rate from our orientation-dependent rate $|G_{00}|^2$ in Fig. \ref{fig:ocs}(b), however, produces a pair of peaks at $\beta=38^\circ$ and $\beta=142^\circ$ instead. The absence of a maximum at $\beta=90^\circ$ in our result is due to insufficient suppression of the secondary peak at $\beta=28^\circ$ in the orientation-dependent curve of Fig. \ref{fig:ocs}(b) and the global maximum in the same figure occurring at a slightly too large of an angle $\beta$. Despite this discrepancy in the alignment-dependent curve, our orientation-dependent curves of OCS calculated through IR-MEWFAT (Fig. \ref{fig:ocs}(b)) and DFT-based IR-MEWFAT \cite{wahyutama_irmewfat_dft} agree with RT-TDDFT result (see Ref. \cite{wahyutama_irmewfat_dft}). In our opinion, a comparison of orientation-dependent rates, rather than alignment-dependent ones, should offer a higher confidence level since weak structures in the latter tend to get washed out due to the incoherent sum of rates from opposite orientations.} 


\begin{figure}
    \includegraphics[width=\columnwidth]{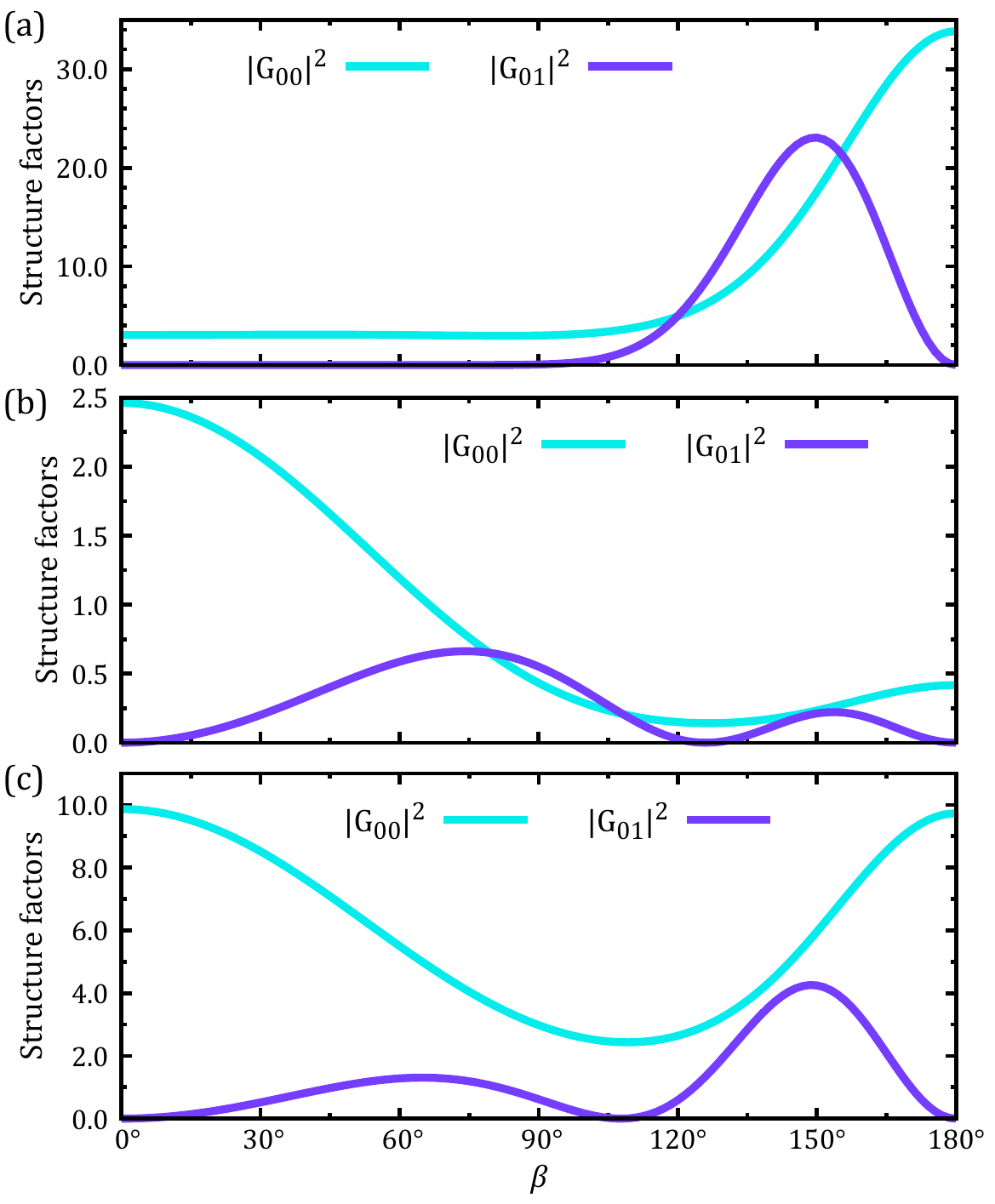}
    \caption{\label{fig:co} Structure factors of CO simulated using (a) OE-WFAT and (b) ME-WFAT. In (a), the HOMO from Hartree-Fock calculation of CO is used. (c) The structure factors calculated with OE-WFAT but where the definitions for the dipole moment and origin use those from ME-WFAT. When $\beta=0^\circ$, the field points from C to O.}
\end{figure}
Next, we turn to  CO whose HOMO has a $\sigma$ symmetry. We choose the C--O bond length to be $1.102$ \AA. The OE-WFAT and ME-WFAT structure factors are shown in Fig. \ref{fig:co}(a) and \ref{fig:co}(b), respectively, and we see a contrasting behavior in the location of the global maximum. IR OE-WFAT predicts that the ionization should be maximum at $\beta=180^\circ$ (as does the TR OE-WFAT in Ref. \cite{wfat_diatomic_collection, wfat2}), that is, when the field points from O to C,
while ME-WFAT predicts the opposite. It is interesting to note that a recent calculation on CO has also been done addressing the effect of the \rlremove{induced multielectron polarization (MEP)} field-induced MEP \cite{multielectron_pol_co} on ionization yields. One of the findings in Ref. \cite{multielectron_pol_co} is that without MEP, the angle-resolved ion yield from CO looks very similar to $|G_{00}|^2$ in Fig. \ref{fig:co}(a), whereas with MEP, 
the ion yield resembles closely $|G_{00}|^2$ of Fig. \ref{fig:co}(b). Experiments \cite{co_experiment} as well as time-dependent Hartree-Fock \cite{co_tdhf} calculations also produce angle-resolved ionization of CO in which the ionization is most favorable when the field points from C to O, and their shape is very similar to the yield with MEP in Ref. \cite{multielectron_pol_co} and $|G_{00}|^2$ in Fig. \ref{fig:co}(b). What is interesting here is that the IR ME-WFAT introduced in this work does not capture MEP effects. \rlremove{(which may be done by including the 1st order correction of ME-WFAT).} 

Repeating the procedure we used for OCS, which mixes the origin and dipole moment of ME-WFAT with the orbital and integral formula of OE-WFAT, we obtain the structure factors in Fig. \ref{fig:co}(c). Unlike the case of OCS for which this procedure results in structure factors that nearly reproduce the ME-WFAT results, the structure factors of both parabolic channels for CO seem to only approximate the corresponding ME-WFAT structure factors in Fig. \ref{fig:co}(b). We attribute 
this disagreement to the non-negligible contribution of the next largest element of $\mathcal P(i)$ after $\mathcal P(\textrm{HOMO})$, which is $\mathcal{P(\textrm{HOMO-2})}$ with $r = 0.114$ (see Table \ref{tab:properties}). For comparison, for OCS this ratio is $0.050$. This observation, along with the one in the previous paragraph, are hints that, in addition to the inclusion of \rladd{field-induced} MEP effects (which requires the first order correction of WFAT), the correct treatment of the permanent dipole moment, as is already captured by ME-WFAT in the leading order approximation, and an accurate representation of the ionizing orbital (in ME-WFAT, the Dyson orbital) are the missing features in the methods that fail to reproduce the experimental data.

We would like to emphasize that the marked change of the structure factors when the origin and dipole moment are changed, \textit{e.g.} from Fig. \ref{fig:co}(a) to Fig. \ref{fig:co}(c) or from Fig. \ref{fig:ocs}(a) to Fig. \ref{fig:ocs}(c), should not be viewed as the failure of ME-WFAT to be origin-independent (which is a necessary property of physical quantities such as ionization rate). When one changes the origin, 
the dipole moment $\mu_z^{n'n}$, the monopole term $Z_c/r$ (see the discussion following Eq. \eqref{eq:v1e_define}), and the center of $R_\nu^{(0)}(\eta) \Phi_\nu^{(0)}(\xi,\varphi)$ must also be transformed accordingly.
In the preceding analysis however, instead of using the properly shifted dipole moment, we use the dipole moment value calculated by ME-WFAT. In fact, we have checked that the OE-WFAT structure factors resulting from shifting the origin and also properly shifting the dipole moment according to the origin shift give results very close to the original OE-WFAT results---Fig. \ref{fig:co}(a) for CO and Fig. \ref{fig:ocs}(a) for OCS. The analysis employed in the current section that mixes ME-WFAT's origin and dipole moment with OE-WFAT's orbital and integral formula may therefore be viewed as an improper specification of dipole moment, and an {\it ad hoc} procedure at best. 

\subsection{N$_2$, HCOOH, and CH$_3$F}  \label{sec:n2_hcooh_ch3f}

\begin{figure}
    \includegraphics[width=\columnwidth]{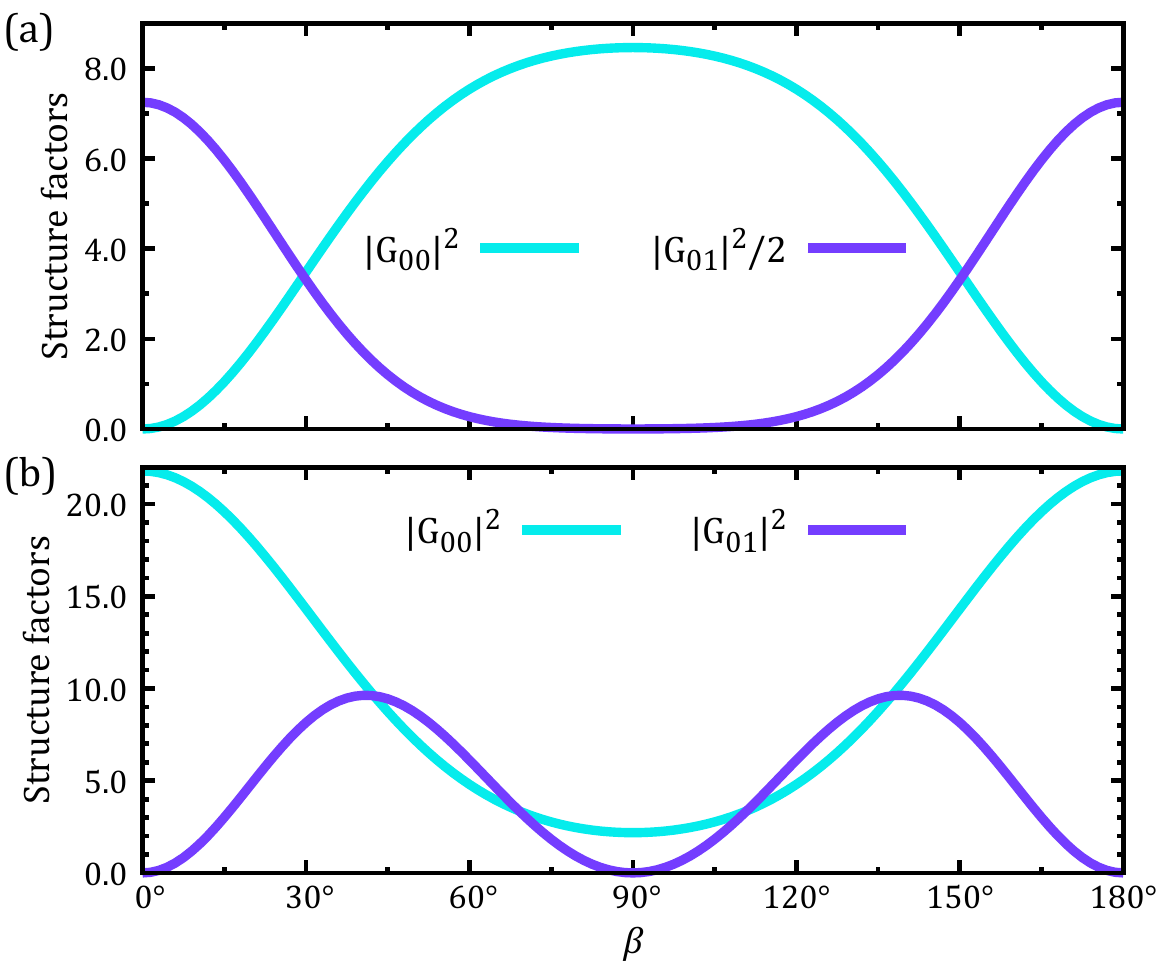}
    \caption{\label{fig:n2} Structure factors of N$_2$ simulated using (a) OE-WFAT and (b) ME-WFAT. In (a), HOMO from Hartree-Fock calculation of N$_2$ is used. The structure factors in (a) are a slice at a constant $\gamma$, which is the direction of the lobe of the HOMO density. When $\beta=0^\circ$, the field is parallel with the molecular axis.}
\end{figure}
In the previous section, we demonstrated that for some molecules, ME-WFAT structure factors differ from the OE-WFAT counterpart mostly 
by the value of the dipole moment used in the exponential factor of Eq. \eqref{eq:f_coeff}. In these cases, ME-WFAT structure factors may be recovered by deliberately using the ME-WFAT dipole moment and origin in the OE-WFAT formulation. We will now show that ME-WFAT and OE-WFAT calculations can differ much less trivially when the Dyson orbital connecting the neutral and cation ground states notably differ from the neutral HOMO.

It is well-known that the occupied MO of N$_2$ calculated by HF have an incorrect 
ordering---experiments suggested that the first ionization from N$_2$ should produce a $\Sigma$ cation state, while Koopman's theorem says that the first ionization should leave the cation in a $\Pi$ state \cite{n2koopman, szabo1996modern, o2_experiment}.
See Ref. \cite{o2_experiment} for an experimental orientation-dependent ionization yield of N$_2$. In the context of OE-WFAT where the ionizing orbital must be chosen manually, the 
HF simulation for N$_2$ can lead to the wrong angle-dependent ionization yield because the $\pi$ character of the false HOMO will obviously dominate since it has the lowest binding energy. 
Using a N--N bond length of $1.066$ \AA, we run OE-WFAT calculation N$_2$ and obtain Fig. \ref{fig:n2}(a) where we see that OE-WFAT $|G_{00}|^2$ calculated using the HOMO of N$_2$ exhibits a $\pi_u$ symmetry. Fig. \ref{fig:n2}(b) shows the results of ME-WFAT, where $|G_{00}|^2$ is seen to have the correct $\sigma$ symmetry. The structure factors shown in Fig. \ref{fig:n2}(b) also look very similar to those obtained by OE-WFAT using HOMO-1 of N$_2$ \cite{wfat2}. 
The reason of the success of ME-WFAT in this case is that the Dyson orbital, being a linear combination of the occupied MO of the neutral, has the largest component (largest $|\mathcal{P}(i)|$) that corresponds to HOMO-1 (see Table \ref{tab:properties}). In a sense, the overlap integral between the neutral and cation states defining the Dyson orbital automatically selects HOMO-1 as the orbital that is closest to the actual ionizing orbital. This result also illustrates how a properly relaxed cation wave function can facilitate the construction of an accurate ionizing orbital.

\begin{figure}[h!]
    \includegraphics[width=\columnwidth]{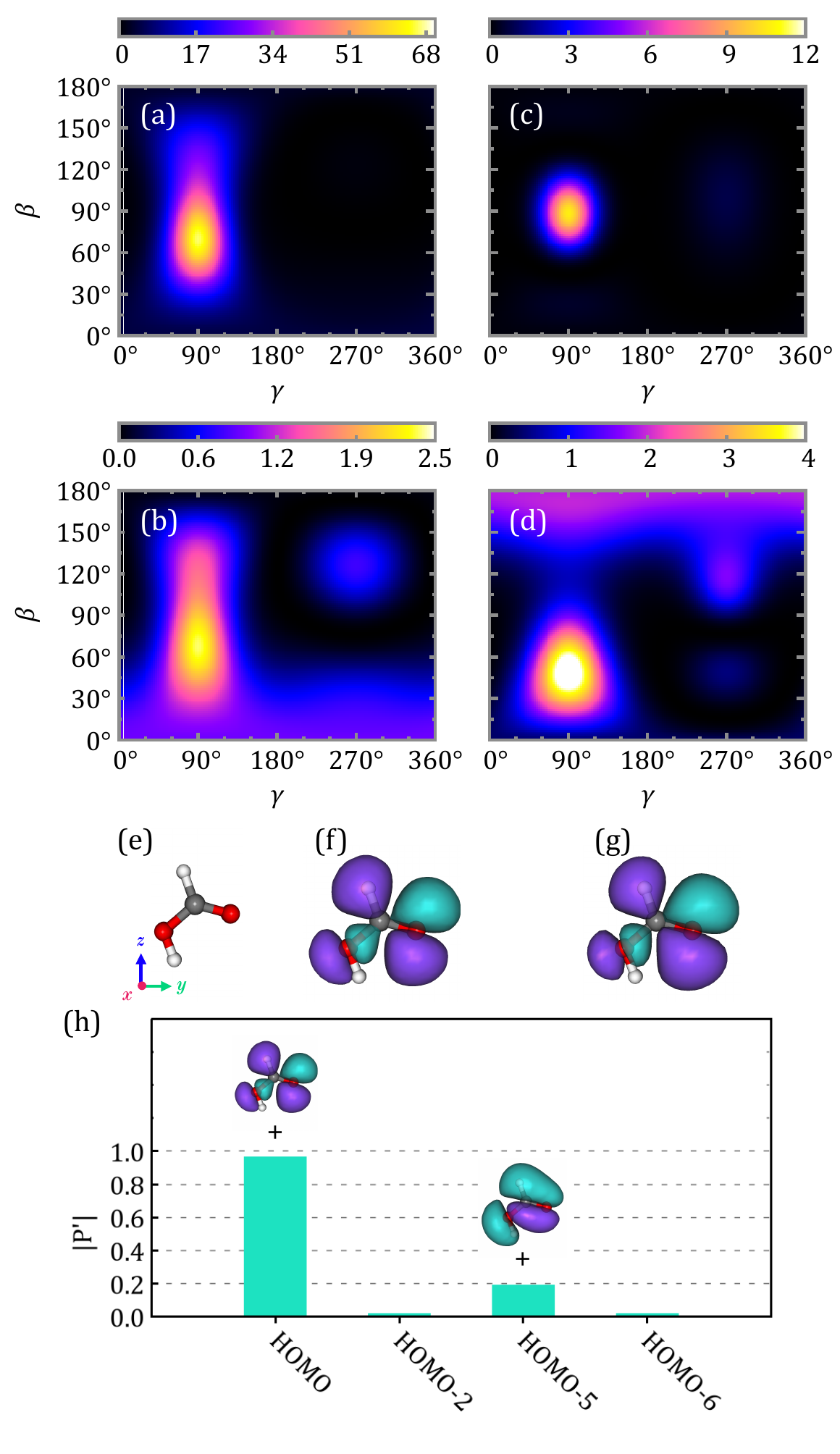}
    \caption{\label{fig:hcooh} 
    \rlremove{The structure factors $|G_{00}|^2$ of HCOOH calculated using (a) OE-WFAT and (b) ME-WFAT. (e) The shape of HCOOH.}
    \rladd{Structure factors $|G_{00}|^2$ from (a) HOMO, (c) HOMO-4, and (d) HOMO-5 of HCOOH calculated using OE-WFAT. The magnitude of $|G_{00}|^2$'s in panel (c) and (d) have been divided by $100$. (b) $|G_{00}|^2$ calculated by ME-WFAT.}
    (e) \rlremove{The shape of HCOOH} \rladd{Image of the HCOOH molecule as well as the $z-$axis which determines the $(0^\circ, 0^\circ)$ direction.} (f) The HOMO used to calculate the $|G_{00}|^2$ in (a). (g) The Dyson orbital used to calculate the $|G_{00}|^2$ in (b).
    (h) The values of $\mathcal P'(i) = (-1)^i \mathcal{P}(i)$ of several MO with the largest contribution to the Dyson orbital. The sign symbol above each bar indicates its sign. The orbital picture shown above the sign symbol is the shape of the corresponding neutral orbital.}
\end{figure}
\begin{figure}[h!]
    \includegraphics[width=\columnwidth]{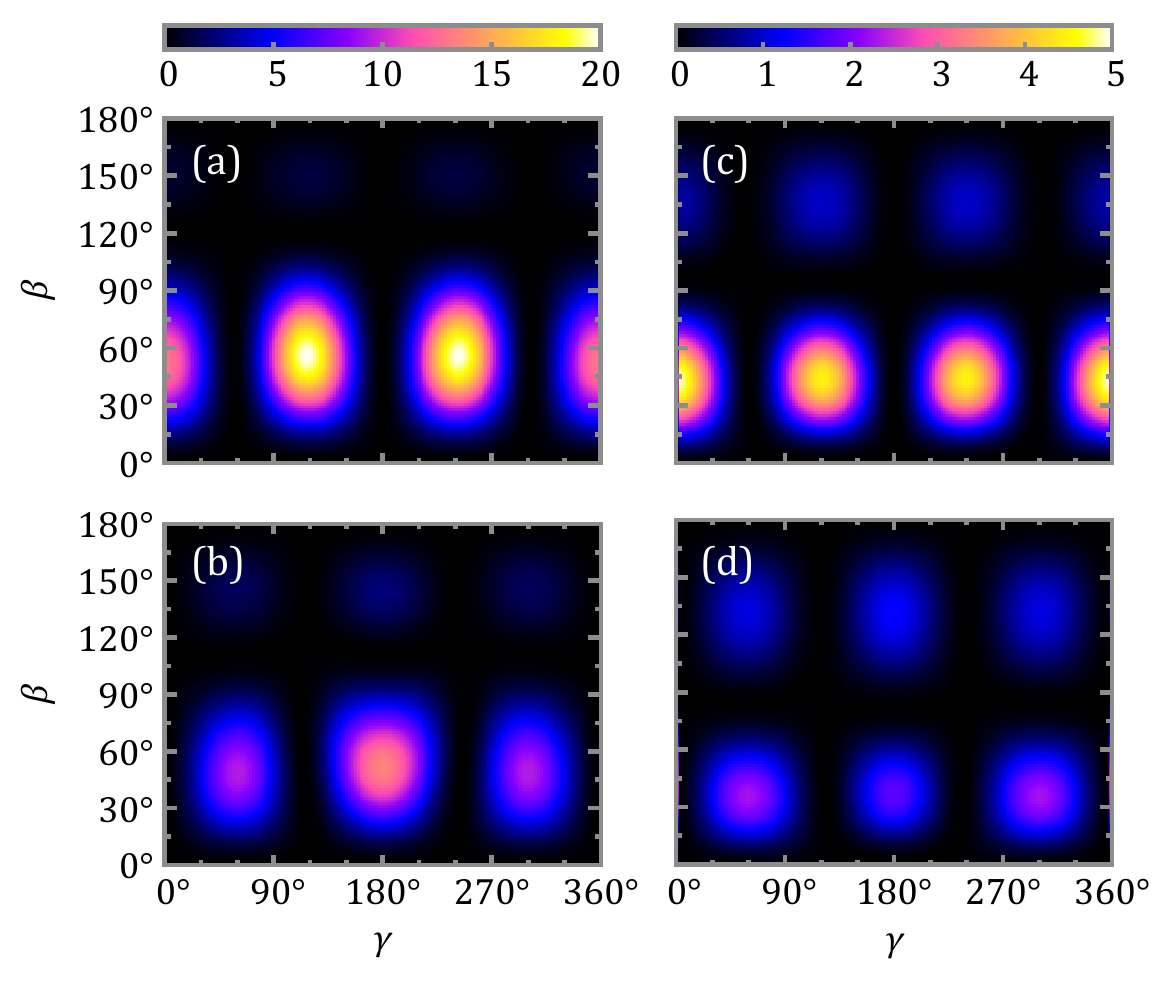}
    \caption{\label{fig:ch3f} (a) and (b): $|G_{00}|^2$ of CH$_3$F obtained by OE-WFAT using \rladdtwo{the two degenerate HOMO of the neutral molecule} \rlremovetwo{HOMO(--) and HOMO(+)}, respectively. (c) and (d): $|G_{00}|^2$ of CH$_3$F obtained by ME-WFAT using \rladdtwo{the two degenerate ground states of the cation} \rlremovetwo{Dyson(--) and Dyson(+)}, respectively. \rladdtwo{These degenerate states are constructed at every orientation angle so that they diagonalize one-particle (in OE-WFAT) or all-particle (in ME-WFAT) dipole matrix.}
    \rlremovetwo{(e) The shape of CH$_3$F. (f) HOMO(--). (g) Dyson(--). (h) The values of $\mathcal{P}'(i)$ of several MO with the largest contribution to Dyson(--).
    The above structure factors are based on the reference that $(\beta,\gamma)=(0^\circ,0^\circ)$ happens when the C--F bond coincides the $z$-axis.}}
\end{figure}
Next, we examine
HCOOH (formic acid), a planar molecule \rladd{(see Fig. \ref{fig:hcooh}(e))}. The atomic coordinates in angstrom when $(\beta,\gamma)=(0^\circ,0^\circ)$ are
C: $(0,  0,      0.415)$,
O: $(0, -1.011, -0.437)$,
O: $(0,  1.139,  0.110)$,
H: $(0, -0.372,  1.444)$, and
H: $(0, -0.658, -1.322)$.
According to Ref. \cite{dyson_orb_hf_orb}, formic acid is a good candidate to study the difference between using the 
Dyson orbital and the HOMO to calculate the ionization rate due to the notable contribution of occupied orbitals other than the HOMO in the linear combination of Eq. \eqref{eq:incomplete_dyson}. $|G_{00}|^2$ resulting from an ionization from the HOMO calculated with OE-WFAT is shown in Fig. \ref{fig:hcooh}(a). It features a single lobe centered around $(\beta,\gamma)=(67^\circ,90^\circ)$.
The ME-WFAT version of this structure factor is shown in Fig. \ref{fig:hcooh}(b), where we see that the single lobe seen in Fig. \ref{fig:hcooh}(a) 
persists.
The most notable difference with OE-WFAT is the appearance of a second, weaker lobe centered around $(\beta,\gamma)=(128^\circ,270^\circ)$ and the 'leakage' of the main lobe toward $\beta=0^\circ$. These features, which are absent in the OE-WFAT structure factor, indicate that there are some occupied orbitals of the neutral that interfere with the dominating HOMO in forming the ionizing orbital. Fig. \ref{fig:hcooh}(f) and \ref{fig:hcooh}(g) compare the shapes of HOMO and Dyson orbital, respectively, with Fig. \ref{fig:hcooh}(h) presenting the individual contribution of some of the most important occupied orbitals of the neutral in terms of their coefficient, $\mathcal{P}(i)$, in the linear combination of Eq. \eqref{eq:incomplete_dyson}. \rlremove{We see that the contribution of HOMO-5 is not so small.}

\rladd{We see that the contribution of HOMO-5 is not so small compared to the largest contribution coming from HOMO. The contribution from HOMO-5 to the Dyson orbital is most likely also responsible for the sharp difference in the magnitude of $|G_{00}|^2$ calculated by OE-WFAT (Fig. \ref{fig:hcooh}(a)) and ME-WFAT (Fig. \ref{fig:hcooh}(b)). This can be seen from OE-WFAT structure factor from HOMO-5 shown in Fig. \ref{fig:hcooh}(d), where we see that it has a maximum at about the same orientation as the structure factor in Fig. \ref{fig:hcooh}(a) 
\footnote{For this analysis, only the angular dependence of OE-WFAT-calculated structure factors from HOMO-4 and HOMO-5 are relevant. The magnitude, however, is irrelevant because it is determined mainly by orbital energies, which are not present in the formulation of ME-WFAT.}. 
This may suggest that the regions of HOMO and HOMO-5 that are responsible for their respective orientation of maximum ionization rate interfere in a way that drastically reduces the maximum ionization probability from the Dyson orbital. There is another MO that produces maximum ionization at about the same angle as HOMO and HOMO-5, namely HOMO-4, whose OE-WFAT structure factor is shown in Fig. \ref{fig:hcooh}(c). However, its corresponding $\mathcal{P}(i)$ is significantly smaller than that of HOMO-5.}


As the last example, we examine
CH$_3$F, a molecule having an equilibrium geometry that possesses a $C_{3v}$ symmetry,
where the axis of the three-fold rotational symmetry coincides with the CF bond.
The C--F and C--H bond lengths are $1.360$ \AA~and $1.080$ \AA, respectively, with the HCF angle of $108.93^\circ$.
Its HOMO is doubly degenerate and can be identified based on their reflection symmetry with respect to \rlremovetwo{the plane containing} one of the three FCH planes.
\rladdtwo{Unlike the degenerate HOMO of OCS and O$_2$, the two degenerate HOMO of CH$_3$F do not diagonalize the dipole matrix, therefore, in view of the perturbation theory, a diagonalization of the dipole matrix in the subspace of degenerate HOMO (in OE-WFAT) or in the subspace of degenerate CH$_3$F$^+$ ground state (in ME-WFAT) must be performed at every orientation angle. The eigenvectors are then used in the WFAT integral formula---for ME-WFAT, these eigenvectors are used as $|\Psi^+\rangle$ wherever it appears. In practice, these dipole-diagonalizing states are never actually constructed at every orientation angle, which would otherwise require the expensive WFAT integrals (Eq. \eqref{eq:mewfat_integral1}) to be recomputed too. This is possible since structure factors are linear with respect to $|\Psi^+\rangle$. Therefore, the transformation can be applied to the structure factors of the non-diagonalizing states, avoiding the recomputation of WFAT integrals.}

\rladdtwo{Fig. \ref{fig:ch3f}(a) and \ref{fig:ch3f}(b) show the $|G_{00}|^2$ structure factors obtained by OE-WFAT using the two degenerate HOMO of CH$_3$F that diagonalize the one-particle dipole matrix. 
While Fig. \ref{fig:ch3f}(c) and \ref{fig:ch3f}(d) are the ME-WFAT counterparts obtained using the two degenerate ground states of CH$_3$F$^+$ that diagonalize the total (all-particle) dipole matrix.
Both OE-WFAT and ME-WFAT structure factors in Fig. \ref{fig:ch3f} exhibit the three-fold rotational symmetry of the molecule. We note in passing that the individual state's structure factors calculated without diagonalizing the dipole matrix do not exhibit the three-fold rotational symmetry but still lead to identical total structure factor as those in Fig. \ref{fig:ch3f}.}

\rladdtwo{Apart from the intensity, the most noticeable difference between the OE-WFAT and ME-WFAT results in Fig. \ref{fig:ch3f} are the sizes and locations of the lobes, with ME-WFAT structure factors having smaller lobes than OE-WFAT and are located at a smaller $\beta$. Upon closer inspection, it can also be seen that the three lobes in the ME-WFAT results are more uniform in shape than those in OE-WFAT. We attribute this observation to the direction of the dipole moment vectors. Due to the approximate nature of the electronic structure computation, the dipole vectors of the degenerate CH$_3$F HOMO (as used by OE-WFAT) are numerically less parallel to the $C_3$ axis of the molecule than the dipole vectors of the neutral's ground state and the cation's degenerate ground states (as used by ME-WFAT).
}

\rlremovetwo{
, here, we denote the symmetric and antisymmetric functions as HOMO(+) and HOMO(--), respectively.
We choose the reflection plane to be the $xz$-plane. Fig. \ref{fig:ch3f}(a) and \ref{fig:ch3f}(b) are the $|G_{00}|^2$ structure factors obtained by OE-WFAT using HOMO(--) and HOMO(+), respectively. The ground state of CH$_3$F$^+$ is also doubly degenerate, which is the consequence of the equal probabilities of removing an electron from either HOMO(+) or HOMO(--) of the neutral to form the ground state cation. This means that one may construct two ``degenerate'' Dyson orbitals, Dyson(+) and Dyson(--). 
The ME-WFAT $|G_{00}|^2$ structure factors using Dyson(--) and Dyson(+) are shown in Fig. \ref{fig:ch3f}(c) and \ref{fig:ch3f}(d), respectively. 
We see that the positions as well as the relative intensities are visibly different compared to OE-WFAT results (Fig. \ref{fig:ch3f}(a) and (b)), for example, the peaks are at smaller $\beta$ for ME-WFAT structure factors (Fig. \ref{fig:ch3f}(c) and \ref{fig:ch3f}(d)) and the middle bottom lobe in Fig. \ref{fig:ch3f}(d) (ME-WFAT) is at a lower intensity than the corresponding lobe in Fig. \ref{fig:ch3f}(b) (OE-WFAT). Again, we can understand this observation by looking at the shapes of the HOMO and Dyson orbital shown in Fig. \ref{fig:ch3f}(f) and (g), respectively, for the odd symmetry. The balancing of the upper and lower pairs of lobes of HOMO(--) (Fig. \ref{fig:ch3f}(f)) in the formation of the Dyson orbital (Fig. \ref{fig:ch3f}(g)) are due to the appreciable interference between HOMO(--) and HOMO-2(--) (HOMO-2 is also doubly degenerate) that has a comparable magnitude of $\mathcal{P}(i)$ (see Fig. \ref{fig:ch3f}(h)). 
}

The strong-field ionization of CH$_3$F has previously been studied using time-dependent configuration interaction singles with a quasi-static field in Ref. \cite{schlegel_ch3x}, where the angle-resolved ionization rate is found to have a peak around $\beta\sim 39^\circ$ for low intensity cases. In this regard, ME-WFAT agrees more closely 
with $\beta=42^\circ$ (see Fig. \ref{fig:ch3f}(c) and \ref{fig:ch3f}(d)) than OE-WFAT with $\beta=52^\circ$ (see Fig. \ref{fig:ch3f}(a) and \ref{fig:ch3f}(b)).

\rladd{\rlremovetwo{As a final note regarding the Dyson orbital, the $\mathcal P(i)$ coefficients, which encode the contribution of the neutral molecule's MO to the Dyson orbital, is independent of the orbital energies of these MO. This is worth mentioning because one might be tempted to think that $|\mathcal{P}(i)|$ should die down monotonically as a function of orbital energy difference relative to that of the orbital that dominates the calculated ionization channel (usually the HOMO for ground-state-to-ground-state channels). This is an incorrect view because these contributions only tell us how much of these MO is needed to build an optimal ionizing orbital (the Dyson orbital). The correct way to compare ionization probabilities
of a group of MO in terms of their orbital energies is by looking at the ionization channels that can be approximated by the removal of an electron from the said MO where their ionization potentials are approximately calculated as an orbital energy of this MO.}}

\section{Conclusions and Outlook} \label{sec:conclusions}

In the present work, we have derived ME-WFAT using the integral representation (IR), allowing for the computation of tunnel ionization in molecules of arbitrary geometries.
This development extends the previous version of ME-WFAT formulated using the tail representation (TR) \cite{tr_mewfat}, where the need for an accurate asymptotic part of the wave functions makes its applicability to systems other than atoms, diatomics, or simple linear molecules difficult. We have presented some test cases covering atoms, as well as linear, planar, and $C_{3v}$-symmetry-possessing molecules to demonstrate its capabilities. The integral formulation outlined in the present work also underlies our recent work on the application of IR ME-WFAT to the DFT wave function \cite{wahyutama_irmewfat_dft}.

We find that the asymptotic coefficients of He, Li, Be, Na, and Mg calculated by IR ME-WFAT generally agree with TR ME-WFAT values with the largest relative error of about $6\%$ for the atoms calculated in this work (see Table \ref{tab:atoms}). The difference can be attributed to the level of theory with which the initial and final wave functions are calculated, which is only at the HF level in this work.
This suggests that the equivalence between TR and IR ME-WFAT should generally improve as more accurate wave functions are used. 

The main factors that can make the angular dependence of the ionization yield calculated by OE-WFAT and ME-WFAT 
noticeably differ 
are (i) the ionizing orbital (the Dyson orbital in ME-WFAT and a MO in OE-WFAT), (ii) the dipole moments, or a combination of both. 
The structure factors calculated by OE-WFAT and ME-WFAT are in very good agreement if the Dyson orbital is close to the HOMO and the HOMO dipole moment is similar to the difference between total initial and final state's dipole moment.
An example of such molecule is O$_2$. 
When the difference in (ii) is more prominent,
new features will appear in ME-WFAT structure factors not seen in OE-WFAT (N$_2$, HCOOH, and CH$_3$F are examples in this category), which include producing a totally different shape for the structure factor. 
From our analysis on N$_2$ (Section \ref{sec:n2_hcooh_ch3f}), we see that ME-WFAT has the property such that it tends to select the most appropriate ionizing orbital having the right symmetry as verified by experiments.

One obvious direction for improvement in the formulation we have presented is to include the first order correction to $f_{\nu\sigma}^{n'n}$ 
to simulate the effect of the distortion of the wave functions on the ionization probability -- this has been done in the TR ME-WFAT \cite{tr_mewfat_first_order}. The inclusion of this induced effect will modify the effective dipole moment of the molecule, which will in turn affect the net force that drives the ionized electron. Another prospect is the use of multiconfiguration wave functions, which has been found to pose a problem for TR ME-WFAT \cite{tr_mewfat_app2}.
In particular, IR ME-WFAT can be used to accurately calculate ionization from molecules dominated by multireference configurations, such as those containing transition metals, provided accurate multireference wave functions are provided.
Apart from the possibility of studying correlation effects in tunnel ionization, ME-WFAT with multiconfiguration wave functions also enables one to ionize from an excited initial state. 
This is expected to become increasingly important as field strength increases as more and more excited levels of the neutral will get populated. 


\begin{acknowledgments}
We thank Oleg Tolstikhin for the enlightening discussion about the origin-invariance property of the WFAT. The author is also grateful to Takeshi Sato for his suggestions regarding the degenerate Hartree-Fock solutions of CH$_3$F$^+$. This work was supported by the U.S. Department of Energy, Office of Science, Office of Basic Energy Sciences, under Award No. DE-SC0012462. Portions of this research were conducted with high performance computational resources provided by Louisiana State University and the Louisiana Optical Network Infrastructure.
\end{acknowledgments}

\appendix

\section{INTEGRAL FORMULAE FOR 
SINGLE-DETERMINANT WAVE FUNCTIONS} \label{sec:HF_integral_derive}

To derive 
integral formulae for  single determinant wave functions, we will first write down the single-determinant wave functions of Eq. \eqref{eq:slater_det} in terms of their respective Laplace expansion containing Slater determinants of $N-2$ electrons,
\begin{align}
    \label{eq:slater_laplace_neutral}
    \Psi(Q_N) 
    =& \,\, 
    \frac{1}{\sqrt{{N(N-1)}}}
    \sum_{\sigma_j} \sum_{j=1}^{N_{\sigma_j}}
    (-1)^{N+\overline j} 
    \psi_j^{\sigma_j}(q)  \nonumber \\
    &\,\,
    \times
    \sum_{\sigma_k} \sum_{k\in \mathcal K_j}
    (-1)^{1+\overline{\overline k}_j}
    \psi_k^{\sigma_k}(q_1)  \nonumber \\
    &\,\,
    \times
    \Psi[\psi_j^{\sigma_j}, \psi_k^{\sigma_k}](Q_{N-2}) 
\end{align}
\begin{align}
    \label{eq:slater_laplace_cation}
    \Psi^+(Q_{N-1}) 
    =& \,\,
    \frac{1}{\sqrt{N-1}}
    \sum_{\sigma_{k'}} \sum_{k'=1}^{N'_{\sigma_{k'}}}
    (-1)^{1+\overline{k'}}
    \upsilon_{k'}^{\sigma_{k'}}(q_1)   \nonumber \\
    &\,\,
    \times
    \Psi^+[\upsilon_{k'}^{\sigma_{k'}}](Q_{N-2}),
\end{align}
where
\begin{subequations}
   \label{eq:HF_index}
   \begin{gather}
       \overline{j} = \delta_{\sigma\beta}N_\alpha + j , \\
       \mathcal{K}_j \equiv 
       \left\{
       \begin{array}{ll}
           \{k | \, k\in\mathbb Z, 1\leq k \leq N_{\sigma_k}\} &,   \sigma_k    \neq \sigma_j , \\
           &\\
           \{k | \, k\in\mathbb Z, 1\leq k \leq N_{\sigma_k}, k\neq j\} &,      \sigma_k=\sigma_j ,
       \end{array}
       \right.  \\
       \overline{\overline k}_j =
       \left\{
       \begin{array}{ll}
           \overline{k}      &, \overline{k} < \overline{j}\\
           \overline{k} - 1  &, \overline{k} > \overline{j}
       \end{array}
       \right. ,\\
       \overline{k} = \delta_{\sigma\beta} N_\alpha + k , \\
       \overline{k'} = \delta_{\sigma\beta}N'_\alpha + k' ,
   \end{gather}
\end{subequations}
and $A[a,b,\ldots,z]$ denotes a Slater determinant formed by removing $a,b,\ldots,z$ spin orbitals from Slater determinant $A$. 
Although it has no physical significance, the assumption introduced in the beginning of Section \ref{sec:HF_application} on the order of the columns of the matrix yielding the Slater determinants is respected in Eqs. \eqref{eq:slater_laplace_neutral} and \eqref{eq:slater_laplace_cation}. Eqs. \eqref{eq:HF_index} are subject to this spin orbital ordering assumption, that is, different ordering will give different signs of each term in the above Laplace expansions. Such a difference in the sign will, however, only affect the overall sign of the asymptotic coefficient, while the ionization rate, which is an observable, will remain unchanged.

Next, we will introduce the following definitions
\begin{subequations}
   \label{eq:det_coeff_general}
   \begin{align}
       \mathcal{P}(i) &= \operatorname{det}\left(S_{\sigma}^{\Psi^+\Psi} \left[    \varnothing \Big| \psi_i^{\sigma} \right] \right) , \\
       \mathcal{R} &= \operatorname{det}\left(S_{p(\sigma)}^{\Psi^+\Psi} \left[    \varnothing \Big| \psi_i^{\sigma} \right] \right) , \\
       \mathcal{Q}(k,i,k') &= \operatorname{det} \left( S_{\sigma}^{\Psi^+\Psi}    \left[ \upsilon_{k'}^{\sigma} \Big| \psi_i^{\sigma}, \psi_k^{\sigma}    \right] \right) 
       (1-\delta_{ik}), \\
       \mathcal{S}(k,k') &= \operatorname{det} \left(    S_{p(\sigma)}^{\Psi^+\Psi} \left[ \upsilon_{k'}^{p(\sigma)} \Big|    \psi_i^{\sigma} , \psi_k^{p(\sigma)} \right] \right)  .
   \end{align}
\end{subequations}
The notation $S_\sigma^{\Psi^+\Psi} \left[ a',b',\ldots | a, b, \ldots \right]$ denotes the spin-$\sigma$ block of the overlap matrix formed between the orbitals occupied in $|\Psi^+\rangle$ removing spin-orbitals $a',b',\ldots$ and the orbitals occupied in $|\Psi\rangle$ removing spin-orbitals $a,b,\ldots$. The number of orbitals removed from the two wave functions has been ensured such that the resulting modified overlap matrix is square. 
\rladd{Note that $\mathcal{R}$ and $\mathcal{S}$ are independent of $\psi_i^\sigma$, hence $\mathcal{R}$ is a scalar.}
We will also rewrite the 2-electron potential term in Eq. \eqref{eq:v2e_define} as
\begin{align*}
    V_{2e}(\mathbf R_N) &= \sum_{i=1}^{N-1} U_{2e}(\mathbf r_i, \mathbf r), \\
    U_{2e}(\mathbf r_i, \mathbf r) &= \frac{1}{|\mathbf r - \mathbf r_i|}.
\end{align*}
The 2-electron contribution to the integral of Eq. \eqref{eq:mewfat_integral1} from the $U_{2e}(\mathbf r_1, \mathbf r)$ term is
\begin{widetext}
\begin{align}
    \left\langle \Psi^+;\Omega_{lm'}^\nu \chi_{\sigma} \left| \hat U_{2e}(\hat{\mathbf r}_1, \hat{\mathbf r}) \right| \Psi \right\rangle
    =& \,\,
    \frac{1}{\sqrt{N}(N-1)}
    \sum_{\sigma_k} \sum_{k'=1}^{N'_{\sigma_k}}
    \sum_{j=1}^{N_\sigma}
    \sum_{k\in\mathcal K_j}
    (-1)^{N+\overline{k'}+\overline{j}+\overline{\overline k}_j}
    \left\langle 
    \Psi^+\left[\upsilon_{k'}^{\sigma_{k}}\right] 
    \big| 
    \Psi\left[\psi_j^\sigma, \psi_k^{\sigma_k}\right] 
    \right\rangle    \nonumber \\ 
    &\,\,
    \times
    \int d^3\mathbf r \,
    (\Omega_{lm'}^\nu(\mathbf r))^*  
    \big\langle \upsilon_{k'}^{\sigma_k} \big| \hat U_{2e}(\hat{\mathbf r}_1, \mathbf r) \big| \psi_k^{\sigma_k} \big\rangle \,
    \psi_j^\sigma(\mathbf r) \nonumber \\
    =& \,\,
    \frac{\delta_{M'_s+m_\sigma, M_s}}{\sqrt{N}(N-1)}
    (-1)^N
    \left\{
    \mathcal R
    \sum_{k'=1}^{N'_\sigma} \sum_{j=1}^{N_\sigma} \sum_{k\neq j}^{N_\sigma}
    (-1)^{\overline{j}+\overline{k'}+\overline{\overline k}}
    \mathcal Q(k,j,k')
    \langle \Omega_{lm'}^\nu | \hat V_{k'k}^\sigma | \psi_j^\sigma \rangle
    \right. \nonumber \\
    &\,\,
    +
    \left.
    \left\langle \Omega_{lm'}^\nu \right|
    \left(
    \sum_{k'=1}^{N_{p(\sigma)}}  \sum_{k=1}^{N_{p(\sigma)}} 
    (-1)^{\overline{k'}+\overline{\overline k}}
    \mathcal{S}(k',k)
    \hat V_{k'k}^{p(\sigma)}
    \right)
    \left(
    \sum_{j=1}^{N_\sigma}
    (-1)^{\overline{j}}
    \mathcal{P}(j)
    \left|\psi_j^\sigma \right\rangle
    \right)
    \right\} \nonumber \\
    =& \,\,
    \frac{\delta_{M_s' + m_{\sigma}, M_s}}{\sqrt{N}(N-1)} \,    (-1)^{N+\delta_{\sigma\beta}N_\alpha}  \,
       \Bigg\{  
       \mathcal{R} \sum_{k'=1}^{N_{\sigma}-1} \, \sum_{j=2}^{N_{\sigma}} \,    \sum_{k=1}^{j-1} (-1)^{j+k+k'} \mathcal{Q}(k,j,k')  \nonumber \\
    & \,\, 
    \times
    \left( \left\langle \Omega_{lm'}^\nu \left| \hat V_{k'k}^{\sigma} \right|    \psi_j^{\sigma} \right\rangle   
       -\left\langle \Omega_{lm'}^\nu \left| \hat V_{k'j}^{\sigma} \right|    \psi_k^{\sigma} \right\rangle \right) 
       + 
       \big\langle \Omega_{lm'}^\nu \big| \hat{\mathcal{V}}^{\sigma} \big|    \tilde{\psi}^{\sigma} \big\rangle 
       \Bigg\}.
    \label{eq:u2e_term}
\end{align}
\end{widetext}
where
\begin{align}
    \tilde{\psi}^{\sigma} (\mathbf r) &= \sum_{i=1}^{N_{\sigma}} (-1)^i \mathcal{P}(i) \psi_i^{\sigma}(\mathbf r), 
    \label{eq:incomplete_dyson}
\end{align}
\begin{align}
    \mathcal{V}^{\sigma}(\mathbf r) &= \sum_{k'=1}^{N_{p(\sigma)}}  \sum_{k=1}^{N_{p(\sigma)}} (-1)^{k'+k} \mathcal{S}(k',k) V_{k'k}^{p(\sigma)}(\mathbf r), \\
    V_{k'i}^{\sigma} (\mathbf r) &= \int d^3\mathbf r' \left(\upsilon_{k'}^{\sigma} (\mathbf r') \right)^*  \frac{1}{|\mathbf r - \mathbf r'|} \, \psi_i^{\sigma}(\mathbf r').
\end{align}
In the second equality in Eq. \eqref{eq:u2e_term}, we use the relation that the overlap between two Slater determinants is equal to the determinant of the overlap matrix formed by the orbitals making up each determinant,
\begin{align*}
    \big\langle 
    \Psi^+ & \left[\upsilon_{k'}^{\sigma_{k}}\right] 
    \big|  
    \Psi\left[\psi_j^\sigma, \psi_k^{\sigma_k}\right] 
    \big\rangle \\ 
    =& \,\, 
    \operatorname{det}\left(
    S_{p(\sigma_k)}^{\Psi^+\Psi}\left[\varnothing \big| \psi_j^\sigma\right]
    \right) 
    \operatorname{det}\left(
    S_{\sigma_k}^{\Psi^+\Psi}\left[\upsilon_{k'}^{\sigma_{k}} \big| \psi_j^\sigma, \psi_k^{\sigma_k}\right]
    \right) \nonumber \\
    =& \,\,
    \left\{
    \begin{array}{ll}
        \mathcal R \, \mathcal{Q}(k,j,k')   &, \sigma_k = \sigma \\
        & \\
        \mathcal P(j) \, \mathcal S(k',k)   &, \sigma_k \neq \sigma
    \end{array}
    \right.  ,
\end{align*}
while in the last equality of the same equation, we use the symmetry property of $\mathcal{Q}$, which is $\mathcal Q(k,j,k') = \mathcal Q(j,k,k')$. In this step, we have also expanded 
$\overline{j}$,
$\overline{k'}$, and
$\overline{\overline k}$ based on whether $\sigma_k=\sigma$ or $\sigma_k=p(\sigma)$.
Because the last expression of Eq. \eqref{eq:u2e_term} is independent of electron index, the total 2-electron contribution is then
\begin{align*}
    \left\langle \Psi^+;\Omega_{lm'}^\nu \chi_{\sigma} \left| \hat V_{2e} \right| \Psi \right\rangle 
    =&\,\,
    (N-1)  \nonumber \\
    &\,\,
    \times
    \left\langle \Psi^+;\Omega_{lm'}^\nu \chi_{\sigma} \left| \hat U_{2e}(\hat{\mathbf r}_1, \hat{\mathbf r}) \right| \Psi \right\rangle .
\end{align*}

The 1-electron contribution is much more straightforward to derive,
\begin{align}
    \label{eq:v1e_derive}
    \Big\langle \Psi^+;\Omega_{lm'}^\nu &\chi_{\sigma} \Big| \hat V_{1e}  \Big| \Psi \Big\rangle  \nonumber\\ 
    =& \,\,
    \frac{1}{\sqrt{N}}
    \sum_{j=1}^{N_\sigma}
    (-1)^{N+\overline{j}}  
    \left\langle 
    \Psi^+
    \big| 
    \Psi\left[\psi_j^\sigma\right] 
    \right\rangle
    \langle
    \Omega_{lm'}^\nu | \hat V_{1e} | \psi_j^\sigma 
    \rangle  \nonumber\\
    =& \,\,
    \frac{(-1)^{N+\delta_{\sigma\beta}N_\alpha}}{\sqrt{N}}
    \sum_{j=1}^{N_\sigma}
    (-1)^j  
    \mathcal{R} \, \mathcal{P}(j)
    \langle
    \Omega_{lm'}^\nu | \hat V_{1e} | \psi_j^\sigma 
    \rangle  \nonumber\\
    =&\,\,
    \delta_{M_s'+m_{\sigma},M_s} \left\langle \Omega_{lm'}^\nu \, \Big| \hat V_{1e} \Big| \psi_D^{\sigma} \right\rangle ,
\end{align}
where
\begin{align}
    \label{eq:dyson_define}
    \psi_D^{\sigma} (\mathbf r) 
    = 
    \frac{(-1)^{N+\delta_{\sigma\beta}N_\alpha}}{\sqrt{N}} \mathcal{R} \, \Tilde{\psi}^{\sigma} (\mathbf r).
\end{align}
$\psi_D^{\sigma}$ is the Dyson orbital corresponding to the removal of an electron from the spin-$\sigma$ channel of the initial HF wave function. The appearance of $\delta_{M'_s+m_\sigma,M_s}$ in Eq. \eqref{eq:u2e_term} and \eqref{eq:v1e_derive} accounts for the fact that Slater determinants are eigenstates of the $z$ projection of the total spin angular momentum operator, and that $\hat V_{2e}$ and $\hat V_{1e}$ do not depend on any spin operators.

\rladdtwo{We note that the $\mathcal P(i)$'s encode the contribution of orbitals to the ionization rate through the Dyson orbital (see Eqs. \eqref{eq:dyson_define} and \eqref{eq:incomplete_dyson}).
On the other hand, orbital energy is also often used as a measure of contribution of that orbital to the rate. 
Nevertheless, orbital energy and $\mathcal P(i)$ are independent of each other. But they independently determine the overall contribution of an orbital to the rate---the energies affect the asymptotic behavior through the exponential tail, while the $\mathcal P(i)$'s determine the shape of the ionizing orbital, that is, the Dyson orbital.}


\section{OBTAINING OE-WFAT INTEGRAL FORMULA FROM ME-WFAT}  \label{sec:oewfat_from_mewfat}

Using Eq.  \eqref{eq:det_coeff_unrelaxed} in Eq. \eqref{eq:v2e_term}, we may derive for the 2-electron term
\begin{widetext}
   \begin{align}
       \left\langle \Psi^+;\Omega_{lm'}^\nu \chi_{\sigma} \left| \hat V_{2e}    \right| \Psi \right\rangle 
       =& \,
       \frac{(-1)^{N+\delta_{\sigma2}N_1+i'} }{\sqrt{N}} \,
       \left\langle \Omega_{lm'}^\nu \right|
       \Bigg\{  
       \sum_{k'=1}^{N_{\sigma}-1} \, \sum_{k=1}^{i'-1} (-1)^{k+k'} \delta_{k'k}
       \left(\hat V_{k'k}^{\sigma} \left| \psi_{i'}^{\sigma} \right\rangle
       - 
       \hat V_{k'i'}^{\sigma} \left| \psi_k^{\sigma} \right\rangle \right)  \nonumber \\
       &\, 
       +
       \sum_{k'=1}^{N_{\sigma}-1} \, \sum_{j=i'+1}^{N_{\sigma}} \,  (-1)^{j+k'}    \delta_{k', j-1}
       \left( \hat V_{k'i'}^{\sigma} \left| \psi_j^{\sigma} \right\rangle 
       - 
       \hat V_{k'j}^{\sigma} \left| \psi_{i'}^{\sigma} \right\rangle
       \right) 
       +
       \sum_{k=1}^{N_{p(\sigma)}} \hat V_{kk}^{p(\sigma)}    \left|\psi_{i'}^{\sigma} \right\rangle
       \Bigg\}  \nonumber   \\
       =& \,
       \frac{(-1)^{N+\delta_{\sigma2}N_1+i'} }{\sqrt{N}} \,
       \left\langle \Omega_{lm'}^\nu \right|
       \Bigg\{  
       \sum_{k\neq i'}^{N_{\sigma}}
       \left(\hat V_{kk}^{\sigma} \left| \psi_{i'}^{\sigma} \right\rangle
       - 
       \hat V_{ki'}^{\sigma} \left| \psi_k^{\sigma} \right\rangle \right) + 
       \sum_{k=1}^{N_{p(\sigma)}} \hat V_{kk}^{p(\sigma)}    \left|\psi_{i'}^{\sigma} \right\rangle
       \Bigg\}  \nonumber \\
       =& \,
       \frac{(-1)^{N+\delta_{\sigma2}N_1+i'} }{\sqrt{N}} \,
       \left\langle \Omega_{lm'}^\nu \right|
       \left(  
       \sum_{\sigma_k} \sum_{k=1}^{N_{\sigma_k}}
       \hat V_{kk}^{\sigma_k} \left| \psi_{i'}^{\sigma} \right\rangle
       - 
       \sum_{k=1}^{N_{\sigma}}
       \hat V_{ki'}^{\sigma} \left| \psi_k^{\sigma} \right\rangle 
       \right).
       \label{eq:v2e_term_oe}
   \end{align}
\end{widetext}
In Eq. \eqref{eq:v2e_term_oe}, one may identify the first term in the parentheses to be the classical repulsion term and the second term to be the HF exchange term. While using Eq. \eqref{eq:det_coeff_unrelaxed} in Eq. \eqref{eq:v1e_term} gives the 1-electron term,
\begin{align}
    \label{eq:v1e_term_oe}
    \left\langle \Psi^+; \Omega_{lm'}^\nu \chi_{\sigma} 
    \left| 
    \hat V_{1e}
    \right|
    \Psi \right\rangle 
    =& \,
    \frac{(-1)^{N+\delta_{\sigma2}N_1+i'}}{\sqrt{N}} \nonumber \\
    &
    \times
    \left\langle \Omega_{lm'}^\nu \, \Big| \hat V_{1e} \Big| \psi_{i'}^{\sigma} \right\rangle.
\end{align}
Eq. \eqref{eq:v2e_term_oe} and \eqref{eq:v1e_term_oe}, up to a common constant prefactor, are identical to the integral formula of OE-WFAT in the IR using the occupied orbital $|\psi_{i'}^\sigma  \rangle$ obtained through HF method (compare with eqs. (32)-(34) in Ref. \cite{ir_oewfat_hf}). The sign in the prefactor is immaterial, whereas the $N^{-1/2}$ factor will be canceled by the factor $N$ in Eq. \eqref{eq:ion_rate} upon calculating the ionization rate.


\bibliography{apssamp}

\end{document}